\documentclass[structabstract]{aa}  
\usepackage{graphicx}
\usepackage{txfonts}
\usepackage{natbib}
\usepackage{amsfonts}
\usepackage{color}
\bibpunct{(}{)}{;}{a}{}{,}
\newcommand{\kms}{km\,s$^{-1}$}

\newcommand{\mum}{\,$\mu$m}

\newcommand{\degree}{$^{\circ}$}
\newcommand{\Msun}{M$_{\odot}$}
\newcommand{\Lsun}{L$_{\odot}$}
\newcommand{\Rsun}{R$_{\odot}$}
\newcommand{\Lp}{L\arcmin}
\newcommand{\Ks}{K$_{\rm S}$}
\newcommand{\msunyr}{M$_{\odot}$yr$^{-1}$}
\newcommand{\simil}{$\approx$}
\newcommand{\Av}{$A_{\rm V}$}
\newcommand{\Mdotacc}{$\dot M_{\rm{acc}}$}

\newcommand{\fig}{Fig.}
\newcommand{\figs}{Figs.}
\newcommand{\sek}{Sect.}
\newcommand{\seks}{Sections}
\newcommand{\tab}{Table}

\begin{document}

\title{Variable accretion as a mechanism for brightness variations in T\,Tau\,S\thanks{Based on observations obtained at ESO's La Silla/Paranal observatory under programmes 280.C-5035, 068.D-0537, 070.C-0544, 070.C-0162, 072.C-0593, 074.C-0699, 074.C-0396, 078.C-0386, 380.C-0179, 382.C-0324, and 084.C-0783, as well as observations obtained at UKIRT.}}

\author{R.~van~Boekel\inst{1} 
   \and A.~Juh\'asz\inst{1} 
   \and Th.~Henning\inst{1} 
   \and R.~K\"ohler\inst{2}
   \and Th.~Ratzka\inst{3,4}
   \and T.~Herbst\inst{1,5}
   \and J.~Bouwman\inst{1} 
   \and W.~Kley\inst{6}
} 

\institute{
Max-Planck Institut f\"ur Astronomie, K\"onigstuhl 17, D-69117 Heidelberg, Germany
  \and
Landessternwarte, Zentrum f\"ur Astronomie der Universit\"at Heidelberg, K\"onigstuhl 12, D-69117 Heidelberg, Germany
  \and
Universit\"ats-Sternwarte M\"unchen, Scheinerstra\ss e 1, D-81679 M\"unchen, Germany
  \and
Astrophysikalisches Institut Potsdam, An der Sternwarte 16, D-14482 Potsdam, Germany
  \and
Herzberg Institute of Astrophysics, National Research Council of Canada, 5071 West Saanich Rd, Victoria, BC, V9E 2E7 Canada
  \and
Institut f\"ur Astronomie \& Astrophysik, Universit\"at T\"ubingen, Auf der Morgenstelle 10, D-72076 T\"ubingen, Germany
}

   \date{Received 2009 December ; accepted }


\abstract
{The southern "infrared companion" of T\,Tau is known to show strong photometric variations of several magnitudes on timescales of years, as well as more modest $\lesssim$1\,mag variations on timescales as short as one week. The physical mechanism driving these variations is debated, intrinsic luminosity variations due to a variable accretion rate were initially proposed, but later challenged in favor of apparent fluctuations due to time-variable foreground extinction.}
{We seek to investigate the nature of the observed photometric variability. Based on simple geometric arguments and basic physics laws, a minimum variability timescale can be derived for which variable extinction is a viable mechanism. Because this timescale increases rapidly with wavelength, observations at long wavelengths provide the strongest constraints.}
{We used VISIR at the VLT to image the T\,Tau system at two epochs in February 2008, separated by 3.94~days. In addition we compiled an extensive set of near- and mid-infrared photometric data from the literature, supplemented by a number of previously unpublished measurements, and constructed light curves for the various system components. We constructed a 2D radiative transfer model for the disk of T\,Tau\,Sa, consisting of a passively irradiated dusty outer part and a central, actively accreting component.}
{Our VISIR data reveal a $+$26$\pm$2\% change in the T\,Tau\,S/T\,Tau\,N flux ratio at 12.8\,$\mu$m within four days, which can be attributed to a brightening of T\,Tau\,Sa. Variable extinction can be excluded as a viable mechanism for the observed flux variation based on the short timescale and the long observing wavelength. We show that also the large long-term photometric variability and its associated color-magnitude behavior can be plausibly explained with variable accretion. However, variable extinction is also a viable mechanism for the long-term variability, and a combination of both mechanisms may be required to explain the collective photometric variability of Sa.}
{We conclude that the observed short-term variability is caused by a variable accretion luminosity in T\,Tau\,Sa, which leads to substantial fluctuations in the irradiation of the disk surface and thus induces rapid variations in the disk surface temperature and IR brightness. Both variable accretion and variable foreground extinction can plausibly explain the long-term color and brightness variations. We suggest that the periods of high and variable brightness of Sa that we witnessed in the early and late 1990s were due to enhanced accretion induced by the periastron passage of Sb, which gravitationally perturbed the Sa disk.}


\maketitle

\section{Introduction}

Six decades ago \cite{ambartsumian1947,ambartsumian1949} first proposed T\,Tauri stars to be newly formed low-mass stars. They had been defined earlier as a class of optically variable stars with emission lines in the vicinity of bright or dark nebulosity by \cite{1945ApJ...102..168J}. Ever since, the name-giving object of the class, T\,Tau, has been regarded as the proto-type of young low-mass stars. It is a K0 star with a mass of approximately 2\,M$_{\odot}$ and an age of 1-2\,Myr, that has a modest and time-variable line of sight extinction of \Av\simil1\,mag \citep{1974A&AS...15...47K,2005AstL...31..427M,2007ApJ...671..546L}. It resides in the Taurus molecular cloud at a distance of 148\,pc \citep{2007ApJ...671..546L}.

An "infrared companion", discovered by \cite{1982ApJ...255L.103D}, is located approximately 0\farcs7 south of T\,Tau. It was soon realized that the bolometric luminosity of this source (T\,Tau\,S) rivals that of the optically visible star (T\,Tau\,N), and that it must be a self-luminous object rather than a dust cloud heated by the northern component. Even though T\,Tau\,S is at least 10\,mag fainter than the northern component in the optical \citep{1998ApJ...508..736S}, it dominates the system flux at infrared wavelengths beyond \simil3\,\mum \ \citep{1991AJ....102.2066G,1997AJ....114..744H,2004ApJ...614..235B}, but it contributes only a minor fraction to the total flux at millimeter wavelengths \citep[][]{1997ApJ...490L..99H,1998ApJ...505..358A}.
The southern component was itself found to be a close binary with components Sa and Sb at a projected separation of \simil50\,mas at the time of discovery \citep{2000ApJ...531L.147K}, corresponding to only 7$-$8\,AU at the system distance.

T\,Tau\,S was shown to be strongly variable at near- and mid infrared wavelengths by \cite{1991AJ....102.2066G}, who detected a \simil2\,mag increase in brightness at 2.2\,\mum \ within $\sim$1\,yr, and a similarly large increase at 10\,\mum \ between two observations separated by about five years. They attributed the brightness fluctuations to variable accretion and derived an accretion rate of 3.6\,$\times$\,$10^{-6}$\,\msunyr.

\cite{2004ApJ...614..235B} presented a large set of (near-) simultaneous K and \Lp \ photometry obtained over the course of about seven years starting 1995, and found T\,Tau\,S to vary with a total amplitude of nearly 3\,mag at K-band and more than 2\,mag at \Lp \ during this period. The fastest brightness variation they detected occurred in December 1997, when T\,Tau\,S brightened by 0.9\,mag in the K~band over only a seven day period. They observed the system during three consecutive nights in September 1998 and another three nights in November 1999 in order to detect even faster variations, but did not detect any. The brightness variations in T\,Tau\,S were shown to be color-dependent in a K-\Lp \ vs. K color-magnitude diagram, in a "redder when fainter" fashion that closely resembles the ISM extinction curve \citep{1985ApJ...288..618R}. This behavior was not expected if the observed brightness fluctuations were due to variable accretion, in this case models predicted the opposite behavior, i.e. "bluer when fainter" \citep{1997ApJ...481..912C}, mostly because of the changing contrast ratio between the blue stellar spectrum and the much redder disk spectrum as the accretion rate varies. Therefore \cite{2004ApJ...614..235B} concluded that variable obscuration is the main responsible factor for the observed large brightness variations in T\,Tau\,S, although these authors did also find evidence for variable accretion in T\,Tau\,S \citep{2001ApJ...551.1031B}.

A decade of spatially resolved observations of the southern binary pair have now covered a substantial fraction of the Sa-Sb orbit and have allowed direct determination of the orbital parameters \citep{2000ApJ...531L.147K,2003ApJ...596L..87F,2004ApJ...614..235B,2006A&A...457L...9D,2008A&A...482..929K}. Current estimates still allow a substantial range for some of the orbital parameters, of which the orbital period and the mass of Sb are the most relevant for this paper. Continued observations over the coming years are expected to yield a converged solution. The mass of the primary Sa is \simil2.2\,\Msun, the most recent estimates for the mass of Sb range from 0.4 to 0.8~\Msun, and for the orbital period from \simil25 to \simil100 years \citep{2008A&A...482..929K,2008JPhCS.131a2028K}. The spatially resolved observations have also revealed Sb to show only minor brightness variations at 2.2\,\mum \ ($RMS$\simil0.2\,mag), while Sa has varied by more than 3\,mag during the last decade at this wavelength.

\vspace{0.1cm}
Here we revisit the physical mechanism causing the observed photometric variations of T\,Tau\,S. This work was motivated by short-term mid-infrared variability that was serendipitously detected during an observing campaign with VLT/VISIR in February 2008, primarily aimed at characterizing the [Ne\,{\sc ii}] emission in the T\,Tau system. We reported on the [Ne\,{\sc ii}] observations in a separate paper \citep{2009A&A...497..137V} and here focus on the \simil26\% increase in continuum brightness of T\,Tau\,S at 12.8\,\mum \ that we detected in two observations separated by only four days. In this paper we will distinguish between the very fast variations with modest amplitude as detected in our VISIR observations, and the long-term variations of much larger amplitude that are seen in the entire body of photometry taken over the past decades. These are not necessarily caused by the same physical mechanism.

The paper is organized as follows. In \sek~\ref{sec:observations} we describe the VISIR observations as well as additional previously unpublished infrared measurements performed with a range of facilities. The results of the infrared photometry are reported in \sek~\ref{sec:results}, in which we also construct updated near- and mid-infrared light curves of the various components of the system. 
In \sek~\ref{sec:variable_extinction} we explain why the short-term variability revealed in the VISIR observations cannot be due to variable extinction and must instead be due to intrinsic luminosity variations, i.e. variable accretion. In \sek~\ref{sec:radiativeTransfer} we address the nature of the long-term variability, which has previously been attributed to time-variable extinction. We present radiative transfer calculations of the disk of Sa that include variable accretion, and show that these models qualitatively reproduce the observed long-term color-dependent near- and mid infrared brightness variations. In \sek~\ref{sec:TTauS_scenario} we outline a tentative scenario for T\,Tau\,S in which the periods of high and variable brightness in which Sa has prevailed from \simil1990 to \simil2003 were due to enhanced accretion activity, quantitatively intermediate between an EXOR and a FUOR outburst, induced by gravitational perturbation of the Sa disk during the periastron passage of Sb in \simil1995. We summarize our results in \sek~\ref{sec:summary}.

\section{Observations and data reduction}
\label{sec:observations}
The primary data set that motivated this paper comprises two 12.8\,\mum \ images of the T\,Tau system taken with VISIR at the VLT in February 2008. In addition, we have gathered an extensive set of photometric measurements from the literature, and present previously unpublished mid-infrared data obtained with the MAX instrument at UKIRT and the TIMMI2 instrument at the ESO 3.6m. We also present photometry of all K-band observations performed with NACO at the VLT since 2001, including a number of previously unpublished data sets.

\subsection{VLT VISIR imaging at 12.8\,\mum}
\label{sec:visir_observations}

T\,Tau was observed with the mid infrared imager and spectrograph \emph{VISIR} \citep{2004Msngr.117...12L}, mounted on \emph{Melipal}, the third of VLT's four 8.2\,m Unit Telescopes. Images were taken through an approximately 0.23\,\mum-wide filter centered on 12.81\,\mum. The filter is centered on the [Ne\,II] 12.81\,\mum \ line, but the emission is dominated by the dust continuum and the neon line contributes only \simil5\%  to the total flux integrated over the filter bandwidth. Observations were performed at two epochs, during the nights starting on 2008 February 3 and 7 (see \tab~\ref{tab:visir_observations}).

Standard chopping and nodding techniques were applied to deal with the high atmospheric and instrumental background inherent to ground-based thermal IR observations. In a plain stack of the data the image quality of the first epoch is substantially worse than that of the second epoch. Closer inspection of the data cube\footnote{The T\,Tau system is very bright, allowing the beam shape in each individual frame to be assessed. Both T\,Tau\,N and T\,Tau\,S were detected at $>$100\,$\sigma$ in every frame.} shows that the position of the sources varied by several pixels between different chop half-cycles (here referred to as ``frames''). In a sub-set of the individual frames the beams are strongly distorted, indicating that either the telescope or the chopping secondary mirror has moved \emph{during} the integration. The remaining frames have the same beam quality as the data of the second epoch, where this problem was not encountered. Only frames with sharp, round beams were selected and combined after appropriate alignment. We combined the 25\% best frames into our final images used for the analysis, and verified that our results are independent of this fraction as long as it is below \simil60\%. This reflects the image statistics of the first epoch, in which roughly 40\% of the frames showed distorted beams. This gives us confidence that our method does not have unintended side effects on the photometry performed on the final image as long as all bad frames are removed. We applied the same procedure to the data of the first and second epochs.

In order to determine the system sensitivity and the atmospheric transparency, photometric standards were observed immediately following the T\,Tau observations during both nights. To correct for the airmass difference between the science observations and the calibration measurements, synthetic atmospheric transmission profiles calculated with ATRAN \citep{lord92} were integrated over the filter's spectral response and appropriate correction factors were determined. Because the used narrow band NeII filter lies in a relatively "clean" region of the atmospheric transmission curve, the resulting airmass corrections are small ($\lesssim$6\% during the first night, \simil1\% during the second night). Any uncertainty in these corrections affects the absolute flux calibration but not the relative photometry between T\,Tau\,N and T\,Tau\,S.

Fluxes of the individual components were obtained by performing PSF photometry, using the calibrators observed immediately after the science measurements as PSFs. We slightly smoothed the calibrator observed during the first night by convolving it with a Gaussian of $FWHM$=0\farcs2 to correct for the relatively large difference in airmass between the science and calibration observation.

The accuracy of our \emph{absolute} photometric calibration is dominated by systematic effects, and is difficult to assess because we observed only one calibrator during each night. Thus, the stability of atmospheric transmission was not monitored, and to correct for differences in airmass between the science and calibration observations we had to revert to theoretical calculations, as mentioned earlier. We adopted an uncertainty of 10\% in the absolute calibration of our photometry. We emphasize that the \emph{relative} photometry between T\,Tau\,N and T\,Tau\,S, i.e. of the N/S flux ratio, can be derived with a much higher precision of $\lesssim$2\%, independent of absolute calibration uncertainties.

\subsection{Additional infrared observations}
\label{sec:additional_IR_data}

\subsubsection{UKIRT/MAX imaging}
The T\,Tau system was imaged through a number of filters in the wavelength range 4.7$-$20\,\mum, on the nights starting 1995 November 13, 1996 January 14 and 1996 August 27. The data of the latter epoch were presented earlier by \cite{1997AJ....114..744H}. Here we will include the data in the 12.4\,\mum \ filter, closest to our continuum sampling wavelength of 12.8\,\mum, of all three epochs.

Standard chopping and nodding techniques were applied and the reference star $\alpha$\,Tau, used both for flux calibration and as PSF reference, was in all cases observed directly before or after T\,Tau, with similar sky conditions and airmasses. Separate frames from individual chop cycles (typical exposure times of 0.2\,s) were saved, and later combined using shift-and-add procedures to obtain optimal image quality. Fluxes of the individual components T\,Tau\,N and T\,Tau\,S were extracted with PSF photometry. The components of the southern binary Sa-Sb are not spatially separated in these observations. For further details with regard to the observing and analysis procedures we refer to \cite{1997AJ....114..744H}.

\subsubsection{TIMMI2 imaging and spectroscopy}
\label{sec:timmi2_observations}
The T\,Tau system was observed with the TIMMI2 instrument \citep[][]{1998SPIE.3354..865R} mounted at the ESO 3.6\,m telescope at La Silla observatory, Chile, on the nights starting 2002 February 2 and December 24. Longslit spectra with a North-South slit orientation were taken, as well as imaging observations in the N8.9, N9.8, and N11.9 filters. Photometric and spectroscopic standard stars were observed regularly throughout the respective nights. The spectra have been previously published by \cite{2003A&A...412L..43P}, who did not attempt to extract the fluxes from T\,Tau\,N and T\,Tau\,S separately. In this paper, we have re-analyzed the whole data set, focussing on extracting the northern and southern component separately. 

The TIMMI2 images marginally resolve the N-S system ($\lambda/D$$\approx$0\farcs73 at 12.8\,\mum, compared to the $\approx$0\farcs69 N-S separation, and a 0\farcs2 pixel scale). Individual fluxes for T\,Tau\,N and T\,Tau\,S were extracted from the imaging observations with PSF photometry. In order to correct for small differences in the actual point spread function between calibration and science observations, either the science or PSF reference observation was convolved with a Gaussian of $FWHM$$\lesssim$0\farcs3, where the exact value was chosen such that the residuals in the fit were minimized. We kept as many parameters fixed as possible. In particular, we used the known positions of Sa and Sb with respect to N, and allowed only for a variable contribution of Sa and Sb, giving some freedom to the photocenter of T\,Tau\,S as a whole along the Sa-Sb separation. Thus, we have five free parameters in the fit: the ($x$,$y$) position of T\,Tau\,N, multiplicative factors for N and S, and the relative contribution of Sa and Sb to the total flux of T\,Tau\,S.

A similar procedure was applied to the longslit spectra, somewhat complicated because the 0\farcs45 pixel scale in spectroscopic mode undersamples the PSF. We modeled the observed two-dimensional spectra as recorded on the TIMMI2 detector with the minimum number of free parameters. We used an iterative two-stage procedure aimed at minimizing the residuals between the observed and modeled signal. At each wavelength, the profile in the spatial direction was taken to be the sum of two delta functions with a fixed separation of 0\farcs69, convolved with the profile of a calibrator star (stage~1). The free parameters are the amplitudes of both components, and the fit was done for all wavelengths sequentially. Then, in stage~2, the position and tilt of T\,Tau\,N on the TIMMI2 chip was varied, after which stage~1 was repeated, etc., until convergence was reached and the residuals were minimized. In short, we fitted the amplitudes of T\,Tau\,N and T\,Tau\,S at each wavelength separately, while the position of T\,Tau\,N was fitted to all wavelengths simultaneously and the N-S separation was kept fixed.

\subsubsection{VLT/NACO imaging at 2.2\,\mum}
\label{sec:naco_observations}
Adaptive-optics-assisted imaging of the T\,Tau system was performed with NACO, mounted on YEPUN, the fourth of VLT's 8.2m Unit Telescopes, during numerous epochs since late 2001. 

Standard near-infrared data reduction methods were applied to the NACO images. They were sky-subtracted with a median sky image, divided by a flat field, and bad pixels were replaced by the median of the closest good neighbors. Finally, the images were visually inspected for any artifacts or residuals.

We re-analyzed the entire data set, which includes five previously unpublished observations. Relative photometry of all three stars was performed with T\,Tau\,N as the reference. A good model for the PSF is needed, which includes the seeing halo that is due to the imperfect adaptive optics correction. This requires some care, because the southern binary generally lies well within the seeing halo of the much brighter northern component. We constructed the PSF as follows. We first made an azimuthally averaged radial intensity profile $I(R)$ by performing aperture photometry with 128 circular apertures with logarithmically spaced radii, centered on T\,Tau\,N, but with a 60\degree \ sector centered on T\,Tau\,S masked out (the resulting intensity profile is multiplied by 6/5 to correct for the masked out part). We found that at large radii, where only the seeing halo contributes, the intensity profile can be very well approximated with a quadratic fit in log($I(R)$) vs. $R$ space. At radii of more than 22 pixels ($\approx$0\farcs292) we approximated the PSF with this fit. At smaller radii we used the actual image of T\,Tau\,N as our PSF. In this way we have a PSF that includes the inevitable AO artifacts close to the center and has a well approximated seeing halo, which is not contaminated by the southern binary.

Higher-order AO residuals are seen in all NACO images of T\,Tau and other bright sources. The most pronounced of these are four roughly static features at \simil0\farcs51 from T\,Tau\,N located at \simil45, \simil135, \simil225, and \simil315 degrees East of North. The amplitude of these features is approximately 1\% of the peak flux of Sb, and they are not spatially coincident with T\,Tau\,S. Thus our photometry of the Sa-Sb system is not compromised by our approximation of the PSF with an azimuthally averaged fit at large radii, because any departures from the true PSF are negligible compared to the signal of the southern binary.

The actual PSF photometry was performed with the Starfinder program \citep[][]{2000A&AS..147..335D}, which finds point sources and determines their positions and flux multiplication factors in an image, given the PSF. We carefully checked the residuals, i.e. data minus the synthetic images provided by Starfinder, and found that at the position of Sb we get clean residuals, with occasional minor artifacts of $\lesssim$1\% of the Sb peak flux. This shows that Sb is a perfect point source at 2.2\,\mum, at the resolution of NACO. Around Sa, however, we see substantial extended emission. This emission has the tendency to slightly boost the Sa flux in the PSF fitting because the PSF wings try to match the extended component. Consequently, the central peak emission is somewhat overestimated, resulting in slightly negative residuals at the position of the Sa point source. We corrected for this by down-scaling the Sa fluxes, so that the residuals at the fitted position of Sa are non-negative. This required down-scaling the fluxes of Sa by typically 5\% and always less than 10\%. See \sek\,\ref{sec:naco_results} and \fig\,\ref{fig:extended_emission_Sa} for further discussion.

\begin{figure}
\includegraphics[height=8.95cm,angle=90]{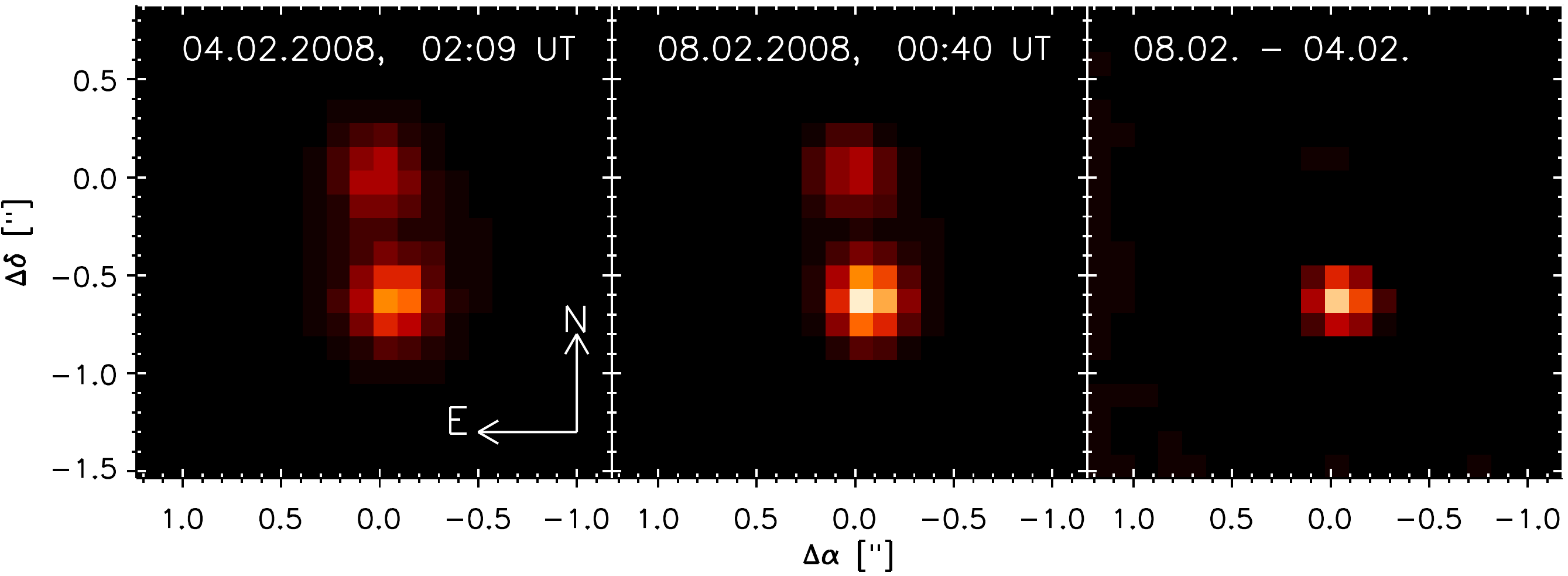}
\caption{\label{fig:image} Two-epoch VISIR imaging of the T\,Tau system in a narrow band filter centered on 12.81\,\mum. The left panel shows the system as it appeared on 2008 February 4, the middle panel as it did four days later, the right panel shows the difference. The fast brightening of T\,Tau\,S is apparent. All images are displayed in a linear stretch chosen in a way that T\,Tau\,N peaks at the same level in both epochs. Values in the difference image have been multiplied by a factor of 4 for better display. Coordinates are in arcseconds relative to T\,Tau\,N. See section~\ref{sec:results} for more details.}
\end{figure}

\begin{table}[b]
\caption{\label{tab:visir_observations} Summary of the VLT/VISIR observations. All images were taken through the [NeII] filter. The last column lists the source brightness at 12.8\,\mum. For the T\,Tau observations, fluxes for the north and south components are individually given. 
The adopted fluxes of the calibrators HD\,41047 and HD\,75691 are taken from \cite{1999AJ....117.1864C}. The absolute flux calibration is good to \simil10\%.}

\begin{tabular}{lcccc}

target  & observing epoch   & airmass & $T_{\rm int}$ & $F_{\nu}$(12.8\,\mum)   \\
        &  (Universal Time) &         &     [s]      &   N / S  [Jy]  \\
\hline

\vspace{-0.2cm}

\\

T Tau     &  2008 Feb 04 \ \ 02:09  &  1.73  & 143.5 &   7.8 / 12.1\\
HD 41047  &  2008 Feb 04 \ \ 02:42  &  1.04  &  54.3 & 5.60        \\
T Tau     &  2008 Feb 08 \ \ 00:40  &  1.45  & 137.8 &   8.5 / 16.7\\
HD 75691  &  2008 Feb 08 \ \ 01:17  &  1.33  &  52.0 & 9.72        \\
\hline
\end{tabular}

\end{table}

\section{Results}
\label{sec:results}

\subsection{Fast photometric variations of T\,Tau\,S at 12.8\,$\mu$m}
\label{sec:fast_variations}

Figure\,\ref{fig:image} shows our 12.8\,\mum \ VISIR images of the T\,Tau system taken during two epochs in early February 2008, separated by 3.94~days. Though T\,Tau shows relatively strong [Ne\,{\sc ii}]  emission at 12.81\,\mum \ \citep{1999A&A...348..877V}, the emission line contributes $\lesssim$5\% to the total system flux integrated over the filter width, and our images are dominated by continuum dust emission. 

The northern and southern components are spatially separated in both images and their fluxes could be determined individually. We measured the flux of T\,Tau\,N to be 7.8 and 8.5~Jy during the first and second night, respectively. We attribute this difference to the uncertainty in our absolute calibration (see also then next paragraph), and in the further discussion we will assume the flux of T\,Tau\,N to be constant at our average value of 8.2\,Jy, consistent with the flux of 8.3\,Jy measured by \cite{2009A&A...502..623R} in December 2004. The fluxes of T\,Tau\,S were correspondingly scaled to 12.8 and 16.1~Jy during the first and second epoch, respectively (note that the values listed in \tab\,\ref{tab:visir_observations} are not scaled but rather correspond to the directly measured fluxes, including absolute calibration uncertainties). 

\subsubsection{How significant is the detected brightness change?}
Taking our measurements at face value, we observed a +9\% (\simil1$\sigma$) flux increase in T\,Tau\,N and a +38\% (\simil4$\sigma$) brightness increase in T\,Tau\,S over the course of four days. The former is clearly not significant, whereas the latter is. The accuracy of these numbers is dominated by uncertainties in the \emph{absolute} photometric calibration. The \emph{relative} photometry between T\,Tau\,N and T\,Tau\,S is much more accurate, we conservatively estimate that we can determine this quantity to an accuracy of 2\%. Thus we can state, as a direct observational result, that the N/S brightness ratio changed by \simil13$\sigma$ ($+$26\%) in four days. Because T\,Tau\,N shows no significant brightness change in our observations, and it is known to be of approximately constant flux in the mid-IR, we have assumed that it did not change in brightness between 2008 February 4 and February 8. Thus, the flux increase in T\,Tau\,S becomes equal to the increase in flux ratio, i.e. +26\% and \simil13$\sigma$. We adopt these values in the current analysis, and note that the assumption of a constant flux level for T\,Tau\,N has no qualitative and only a minor quantitative influence on the reasoning and conclusions presented in this paper.

\subsection{Previously unpublished infrared photometry}
\label{sec:new_midIR_photometry}

\subsubsection{UKIRT/MAX imaging}
\label{sec:max_results}
The measured fluxes of T\,Tau\,N and T\,Tau\,S at 12.4\,\mum, obtained with UKIRT/MAX, are listed in \tab~\ref{tab:max_observations}. Within the uncertainties of the absolute flux calibration, T\,Tau\,N remained at a constant brightness whereas T\,Tau\,S varied substantially between the three epochs of MAX observations.

\begin{table}
\caption{\label{tab:max_observations} UKIRT/MAX photometry of the T\,Tau system at 12.4\,\mum.}
\begin{center}
\begin{tabular}{ r @{} c @{} rcc }
\multicolumn{3}{c}{date} & $F_{\nu}$(N) & $F_{\nu}$(S) \\ 
\hline
1995 & \ \ Nov \ \ & 13 &  9.4   & 16.3     \\
1996 & \ \ Jan \ \ & 14 &  9.8   & 19.7     \\
1996 & \ \ Aug \ \ & 27 &  9.7   & 15.8     \\
\end{tabular}
\end{center}
\end{table}

\subsubsection{TIMMI2 imaging and spectroscopy}
\label{sec:timmi2_results}
\begin{figure}
\includegraphics[width=8.95cm,angle=0]{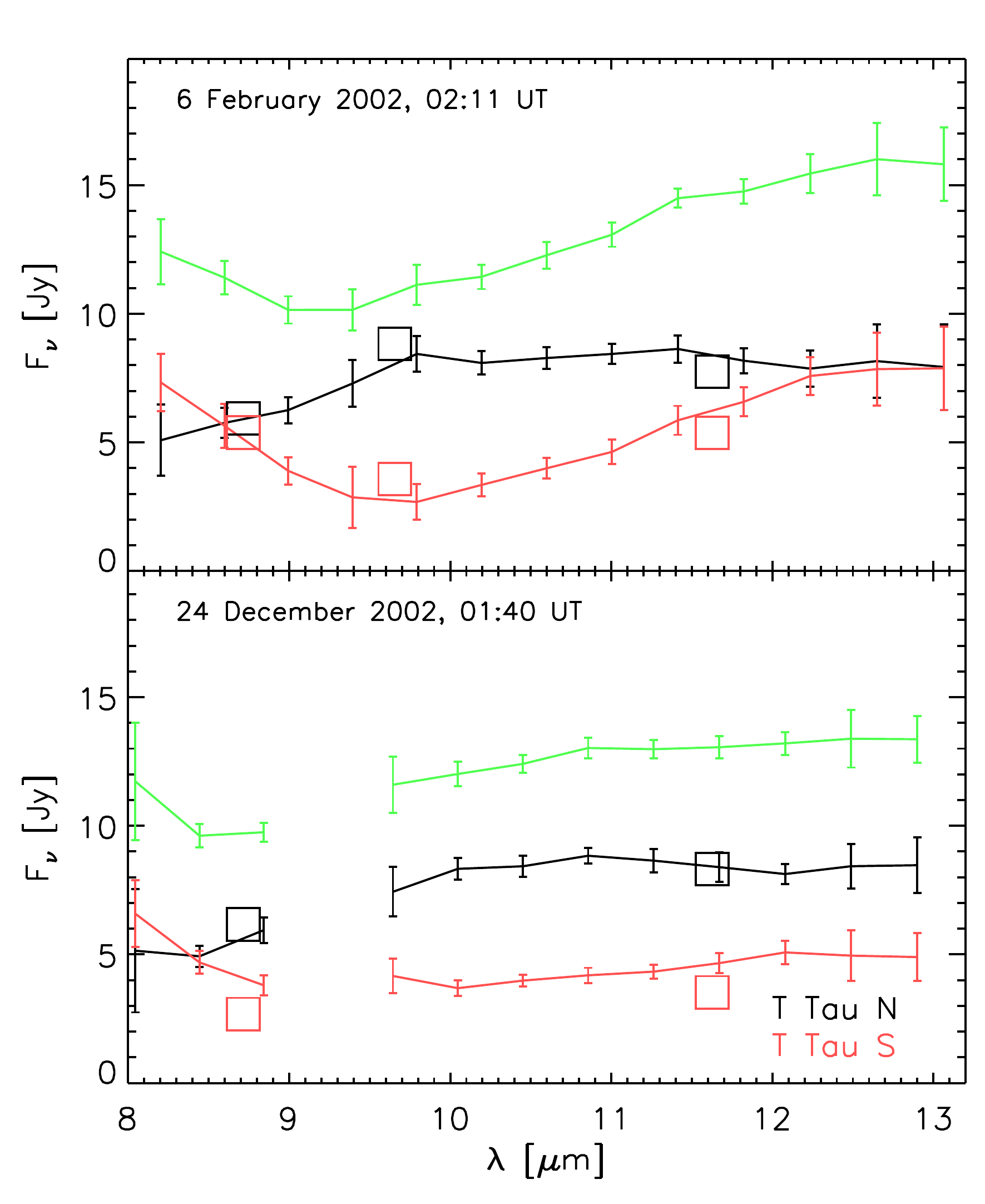}
\caption{\label{fig:timmi2_results} PSF photometry (squares) and PSF spectroscopy performed with TIMMI2 in 2002. The northern and southern components are spatially marginally resolved ($\approx$\,0.9$\lambda/D$ at 12.8\,\mum). See \sek\,\ref{sec:timmi2_observations} for details.}
\end{figure}

Figure\,\ref{fig:timmi2_results} shows the results of our PSF photometry and PSF spectroscopy performed on the 2002 TIMMI2 data. The spectra were rebinned to a resolution of $\approx$25, and the error bars indicate the standard deviation within each bin, i.e. they reflect statistical fluctuations but not necessarily systematic uncertainties in, e.g., the telluric calibration or errors arising in separating the fluxes of both components in the marginally resolved and undersampled data. The ``gap'' in the December spectra around 9.2\,\mum \ is due to a defect detector channel. The spectra taken in February and December 2002 were scaled by factors of 1.05 and 1.29, respectively, to match the flux levels derived from the imaging observations. Such factors may arise from uncertainties in the absolute calibration of the spectra, i.e. due to different slit losses between calibration and science observations. The photometry was not scaled.

The spectra of T\,Tau\,N taken in February and December 2002 are identical in shape within uncertainties and show a weak silicate emission feature that has been reported earlier \citep[e.g.][]{1991AJ....102.2066G,1997AJ....114..744H,2009A&A...502..623R}. In particular the February spectrum of T\,Tau\,S clearly shows the well known silicate absorption feature. The February spectrum shows T\,Tau\,S to be approximately as bright as T\,Tau\,N in the continuum redward of the silicate feature. This spectrum is nearly identical to that measured by \cite{2009A&A...502..623R}. In December the shape of the T\,Tau\,S spectrum has changed substantially: whereas at wavelengths below $\approx$10\,\mum \ it is still very similar to the February spectrum, beyond 10\,\mum \ the red wing of the silicate absorption feature no longer rises stongly, but instead has flattened considerably.

We remind the reader that in our TIMMI2 observations T\,Tau\,N and T\,Tau\,S are only marginally spatially separated, due to the the modest primary mirror diameter of the ESO 3.6m telescope. This limits the accuracy with which the fluxes of the individual components can be extracted. However, with our robust approach using a minimum of free parameters (see \sek\,\ref{sec:timmi2_observations}) we get a fair agreement between the imaging and spectroscopic observations, which is reassuring.

\subsubsection{VLT/NACO imaging at 2.2\,\mum}
\label{sec:naco_results}
In \tab\,\ref{tab:naco_observations} we summarize the VLT/NACO K-band photometry. Most of these data were previously published, but with an emphasis on the relative astrometry  of the various components rather than the photometry \citep[e.g.][]{2008A&A...482..929K,2002Msngr.107....1B}. Here we present the photometry of all existing NACO measurements, including five previously unpublished observations. We did not attempt to perform an absolute photometric calibration. Instead, all photometry of Sa and Sb is relative to T\,Tau\,N. The formal errors on the photometry as given by Starfinder are generally less than 1\% for Sb and less than 2\% for Sa. However, given the non-negligible residuals, in particular surrounding Sa, these estimates seemed too optimistic. We assigned errors of 0.02\,mag to the photometry on Sb and 0.05\,mag on that of Sa. While these estimates may be somewhat conservative, the amplitude of the photometric variations that we are discussing here is \emph{vastly} larger than the uncertainties, and thus our simple error estimates serve the current cause well.

\begin{figure}
\includegraphics[width=8.25cm,angle=90]{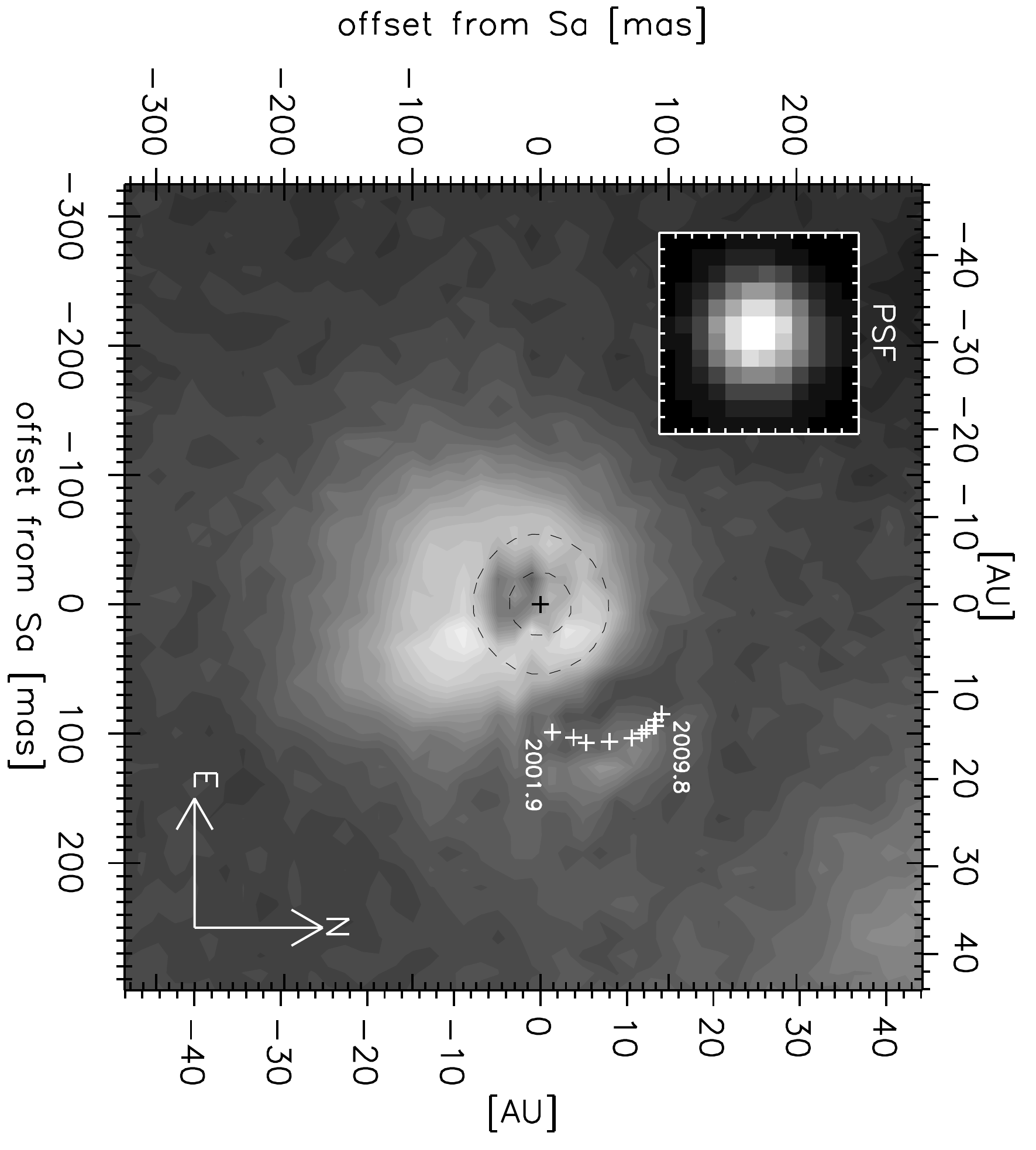}
\caption{\label{fig:extended_emission_Sa} Spatially extended K-band emission around Sa. Shown is a contour plot of the NACO images after subtraction of the best-fit point sources for T\,Tau\,N, Sa, and Sb (see \seks\,\ref{sec:naco_observations} and \ref{sec:naco_results} for details). All epochs were aligned and centered on Sa. $+$ signs mark the positions of Sa and Sb. The inset shows the PSF with the 13.26\,mas pixels of the s13 camera on the same scale as the main plot. The dashed contours show the best-fit point source of Sa, drawn at 25\% and 75\% of the Sa peak flux.}
\end{figure}

As described in \sek\,\ref{sec:naco_observations} we found substantial spatially extended emission around Sa. Around Sb, no such emission was found. In \fig\,\ref{fig:extended_emission_Sa} we show the average residuals of all observations, aligned in a way that Sa is always at the center of the image as indicated with a $+$\,sign. The positions of Sb, one for each epoch, are also indicated with $+$\,signs. The residual emission surrounds Sa, but shows a central cavity. Sa is \emph{not} in the center of this cavity, but rather in the north-west corner. The emission level of the brightest parts of the extended emission corresponds to \simil5\%  of the peak flux of Sb. The emission has its largest extent towards the south. Note that Sa itself has moved by approximately 24\,mas (assuming a mass ratio of 2.2:0.6 for Sa:Sb), or about 1/3 of the diameter of the ``central cavity'' during the period covered by our NACO observations. The spatially extended material we see need not necessarily move along with Sa.

One may ask whether the cavity seen is real or an artefact of imperfect subtraction of the Sa point source. After careful consideration, we conclude that it must be real for the following reaons: (1) the spatially extended emission and the central cavity are persistent over time; (2) the integration times were chosen so that T\,Tau\,N, which is used as PSF reference, stays below the linearity limit of the NACO detector; (3) the cavity is offset from the Sa point source (see the dashed contours in \fig\,\ref{fig:extended_emission_Sa}); and (4) during the observations of the last epoch (October 2009) the field was rotated by $\approx$130$^{\circ}$. This leaves the orientation of PSF on the detector unchanged, but rotates the sky. The extended emission and the central cavity are detected at the same location during this epoch, even though these regions of the sky now correspond to a completely different part of the PSF. This proves that the central cavity cannot be an artifact due to a subtle asymmetry in the PSF.

Our aim here is not to assess the nature of the extended emission in detail. The main point we wish to stress is that \emph{there is material around Sa on scales of \simil5 to 20$-$30\,AU}. This scale is larger than the radius of the Sa disk, which must be $\lesssim$5\,AU, but substantially smaller than the extended H$_2$ 2.12\,\mum \ line emission seen in IFU spectra by \cite{2008ApJ...676..472B} and \cite{2008A&A...488..235G} on scales of \simil50$-$200\,AU. The H$_2$ IFU data do not properly resolve the scales on which we see the extended emisison in the NACO data, but they suggest the H$_2$ emission to be very faint close to Sa. This argues for K-band continuum radiation from Sa scattered by dust as the most likely explanation for the emission we see. Whether the ``central cavity'' is truly devoid of material, or whether there is material present at these locations which has no direct line-of-sight to Sa and the near-IR bright inner regions of its disk and hence does not show up in scattered light, remains to be investigated.

\begin{table}
\caption{\label{tab:naco_observations} Summary of the VLT/NACO photometry in the \Ks \ band obtained between 2001 December and 2009 October. All magnitudes are relative to T\,Tau\,N.}

\begin{center}
\begin{tabular}{ r @{} c @{} rcc }
\multicolumn{3}{c}{date} & $\Delta K_S$ (Sa-N) & $\Delta K_S$ (Sb-N) \\
\hline
 2001 & \ \ Dec \ \ &    8 &   1.67$\pm$0.05  &   3.02$\pm$0.02  \\
 2002 & \ \ Dec \ \ &   15 &   3.61$\pm$0.05  &   3.14$\pm$0.02  \\
 2003 & \ \ Dec \ \ &   12 &   4.75$\pm$0.05  &   3.25$\pm$0.02  \\
 2004 & \ \ Dec \ \ &    9 &   4.79$\pm$0.05  &   3.31$\pm$0.02  \\
 2006 & \ \ Oct \ \ &   11 &   3.14$\pm$0.05  &   2.94$\pm$0.02  \\
 2007 & \ \ Sep \ \ &   16 &   3.09$\pm$0.05  &   2.92$\pm$0.02  \\
 2008 & \ \ Feb \ \ &    1 &   3.07$\pm$0.05  &   3.05$\pm$0.02  \\
 2008 & \ \ Oct \ \ &   17 &   3.46$\pm$0.05  &   3.07$\pm$0.02  \\
 2008 & \ \ Nov \ \ &    6 &   3.84$\pm$0.05  &   2.88$\pm$0.02  \\
 2009 & \ \ Feb \ \ &   18 &   3.45$\pm$0.05  &   2.88$\pm$0.02  \\
 2009 & \ \ Oct \ \ &    7 &   3.05$\pm$0.05  &   3.07$\pm$0.02  \\
\end{tabular}
\end{center}
\end{table}


\newcommand{\yes}{$+$}
\newcommand{\no}{$-$}
\newcommand{\sortof}{$\sim$}

\begin{table}

\caption{\label{tab:IR_photometry} References for infrared photometry used in Figs.~\ref{fig:light_curve}, \ref{fig:KminLp}, and \ref{fig:KminN}. The first three columns give the reference to the original work, the wavelength of observation, and the facility used, respectively. In the last two columns we indicate whether the observations spatially separated the northern and southern component and the Sa-Sb sub-components of T\,Tau\,S, respectively. \tiny{\textcolor{white}{\citeyear{1999A&A...348..877V}}}}

\begin{tabular}{lclcc}
original work & $\lambda_{\rm{obs}}$  & facility &  N-S & Sa-Sb \\

 & [\mum] & & \multicolumn{2}{c}{spat. resolved?} \\

\hline

\vspace{-0.2cm}
 & & & \\

%
\cite{2004ApJ...614..235B} & K, \Lp      & IRTF$^{\dag}$ & \yes & \no$^{\dag}$ \\
\cite{2005ApJ...628..832D} & K, \Lp      & Keck    & \yes & \yes \\
\cite{2007AJ....134..359H}$^{\ddag}$ & \Ks, \Lp      & VLT    & \yes & \yes \\

\cite{1991AJ....102.2066G}  & 10.1, 12.5 & Hale 5m & \yes & \no    \\
Ratzka et al. (2009)        & 12.8 & VLTI      & \yes & \yes   \\
v.d. Ancker et al. (1999)   & 12.8 & ISO       & \no  & \no    \\
this work                   & \Ks  & VLT       & \yes & \yes \\
Ratzka et al. (2009)        & 12.8 & Spitzer   & \no  & \no    \\
this work                   & 12.4 & UKIRT     & \sortof & \no \\
this work                   & 12.8 & ESO 3.6m  & \sortof & \no \\
this work                   & 12.8 & VLT       & \yes & \no    \\

\hline
\end{tabular}
$^{\dag}$\tiny{Includes also data from WIRO, Keck, and Gemini North. The Sa-Sb pair remained unresolved in most of these observations, ut several spatially resolved data points are included as well. Besides a large number of their own measurements, Beck et al. include literature data from \cite{1982ApJ...255L.103D}, \cite{1984ApJ...287..793B}, \cite{1991A&A...249..392M}, \cite{1991AJ....102.2066G}, \cite{1994A&A...283..827T}, \cite{1994PASJ...46L.183K}, \cite{1996ApJ...456L..41S}, \cite{2000IAUS..200P..60R}, \cite{1997AJ....114..744H}, \cite{2001ApJ...556..265W}, \cite{2000ApJ...531L.147K}, \cite{2002ApJ...568..267K}, \cite{2002ApJ...568..771D}, and \cite{2003ApJ...596L..87F}.} \\
$^{\ddag}$\tiny{We applied scaling factors of $\Delta$K=-0.18\,mag and $\Delta$\Lp=+0.32\,mag to all components, so that the magnitudes of T\,Tau\,N match the mean literature values of K=5.52 and \Lp=4.32.
} \\

\end{table}

\subsection{Which components vary by how much?}
The photometric observations of the T\,Tau system performed over the last decades, which form the basis of the current and previous variability studies, had a range of spatial resolutions depending on the facility used and the wavelength of observation. At near-infrared wavelengths, all observations used here spatially resolve T\,Tau\,S from T\,Tau\,N, but only the more recent observations resolve the southern Sa-Sb pair. At mid-infrared wavelengths, most observations spatially resolve T\,Tau\,S from T\,Tau\,N, the space-based measurements yield only the cumulative flux of the whole system, and some measurements spatially resolve the Sa-Sb binary using special techniques.

If one component of the triplet is the dominant source of variability, we can with some care use all measurements to construct a light curve of this source, including those observations that do not spatially separate all three components. We will argue that T\,Tau\,Sa is the main responsible for the photometric variations of the whole system at mid-infrared wavelengths, and that it is dominates the variability of T\,Tau\,S in the near-infrared, as was previously done by e.g. \cite{2005ApJ...628..832D}.

\emph{T\,Tau\,N.} Historic optical photographic plates have revealed that the optically visible component T\,Tau\,N was varying irregularly, rapidly, and strongly between 1858 (beginning of data taking) and $\sim$1917 \citep{lozinskii_1949,2001AJ....122..413B}. Since then, the optical brigthness of T\,Tau\,N has shown comparatively minor ($<$1\,mag) deviations from its average value of B\simil11\,mag, which is approximately the maximum brightness reached during the period of irregular variations prior to 1917, except for brief periods in 1925 and 1931 when the star was substantially fainter. These variations have been attibuted to time-variable line-of-sight extinction caused by dynamic, structured dust clouds passing in front of T\,Tau\,N, possibly related to gravitational interaction between the stars and circumstellar matter in the system, or out-flowing material from T\,Tau\,S \citep{1994AJ....108.1906H,2001AJ....122..413B}. Such variations in foreground extinction may cause infrared variability as well, but the amplitude is expected to be much smaller, in particular in the mid-infrared.

Since the beginning of near- and mid-infrared observations in the 1980s, T\,Tau\,N has not shown substantial variations at these wavelengths \citep[for the near-IR see][]{2004ApJ...614..235B}. At 12.8\,$\mu$m, the fluxes reported in the literature and the new data presented here average around 8.1\,Jy, with a standard deviation of 0.9\,Jy \citep[][]{1991AJ....102.2066G,2008ApJ...676.1082S,2009A&A...502..623R}. Considering the limited accuracy of ground-based absolute photometry in the mid-IR and that these measurements were taken through different filters, requiring some extrapolation to our sampling wavelength of 12.8\,\mum, we conclude that the existing measurements show no evidence for significant mid-IR variability of T\,Tau\,N. Minor variations with an amplitude of $\lesssim$10\% cannot be excluded based on the currently available data.

\emph{T\,Tau\,S.} Near-infrared imaging on AO-assisted 10m class telescopes performed since the early 2000s has revealed Sa to be variable by over 3\,mag at 2.2\,\mum, whereas Sb shows only modest variations with an RMS of 0.2\,mag around a mean value of \Ks\simil8.6 (Note that \cite{2000ApJ...531L.147K} finds a substantially lower flux of K=9.37$\pm$0.25 for Sb in speckle holographic imaging performed in late 1997, assuming T\,Tau\,N has K=5.52). At 12.8\,\mum \ T\,Tau\,S has varied in brightness from \simil5 to \simil27~Jy between the epochs covered \citep[][]{1991AJ....102.2066G}. Recent N-band observations that use mid-IR adaptive optics and interferometric techniques have spatially resolved the Sa-Sb pair in the N-band, and showed that Sb has a brightness of $\lesssim$2.5\,Jy at 12.8\,\mum \ \citep{2008ApJ...676.1082S,2009A&A...502..623R}. Thus, it appears very unlikely that Sb contributes dominantly to the total mid-infrared flux of T\,Tau\,S at any epoch, and thus also not to its variability. This is agrees completely with the absence of large near-infrared variations in Sb. As argued in the previous paragraph, T\,Tau\,N shows no significant IR variability. \emph{Therefore we can reasonably attribute the vast majority of the near-infrared variability of T\,Tau\,S and the mid-infrared variability of the whole T\,Tau system to Sa only}.

\vspace{0.1cm}

We can now deduce the flux of Sa from observations that do not spatially separate all components of the system. In the near-infrared, where all used measurements spatially resolve T\,Tau\,S from T\,Tau\,N, we can obtain the magnitudes of Sa alone by taking Sb to be of constant brightness at K=8.6\,mag \citep[][this work]{2005ApJ...628..832D}. For epochs during which T\,Tau\,S was brighter than K=8.0\,mag, the 0.2\,mag RMS variations of Sb will introduce an error of $\lesssim$0.3\,mag on the thus derived brightness of Sa. Because we are interested in global trends rather than high-precision photometry of Sa, uncertainties of $\lesssim$0.3\,mag are adequate for our purposes (the total variations we are studying amount to $\Delta$K$\gtrsim$3\,mag). We conservatively disregard any near-infrared observations in which the Sa-Sb pair is not spatially separated and the total brightness of T\,Tau\,S is less than K=8.0\,mag in the \cite{2004ApJ...614..235B} data. Likewise, we calculate the \Lp \ brightness of Sa from the total brightness of T\,Tau\,S assuming that Sb has a constant brightness of \Lp=6.25\,mag \citep[][]{2005ApJ...628..832D}. Because Sa is substantially brighter than Sb in \Lp, the uncertainties on the derived \Lp \ magnitudes of Sa will be smaller, though strictly speaking we do not know how variable Sb is in \Lp \ because no multi-epoch, spatially resolved observations in this band are available.

In a similar fashion, we can infer the brightness of Sa at 12.8\mum \ from observations that do not spatially separate Sa from Sb (as is the case for most ground-based observations used here), or from space-based observations that yield only the cumulative flux of the entire T\,Tau system. T\,Tau\,S was much brighter than Sb at all covered epochs, and T\,Tau\,S was approximately as bright as (Spitzer epoch) or substantially brighter than (ISO epoch) T\,Tau\,N in the space-based observations. Therefore the uncertainties in the derived 12.8\mum \ flux for Sa are $\lesssim$0.5\,mag for the epochs during which Sa was faintest and substantially smaller during those epochs when Sa was bright.

In summary, we attribute the vast majority of the observed near-infrared photometric variations in T\,Tau\,S and the mid-infrared variability of the entire T\,Tau system \emph{to Sa only}. We acknowledge the minor near-infrared variations detected in Sb and potential $\lesssim$10\% mid-IR variability in T\,Tau\,N, but these are negligible compared to the Sa variability and have no qualitative and only a very minor quantitative influence on our discussion.

\subsection{Updated light curves of the T\,Tau system}
\label{sec:update_light_curves}

\begin{figure*}
\includegraphics[width=10.0cm,angle=90]{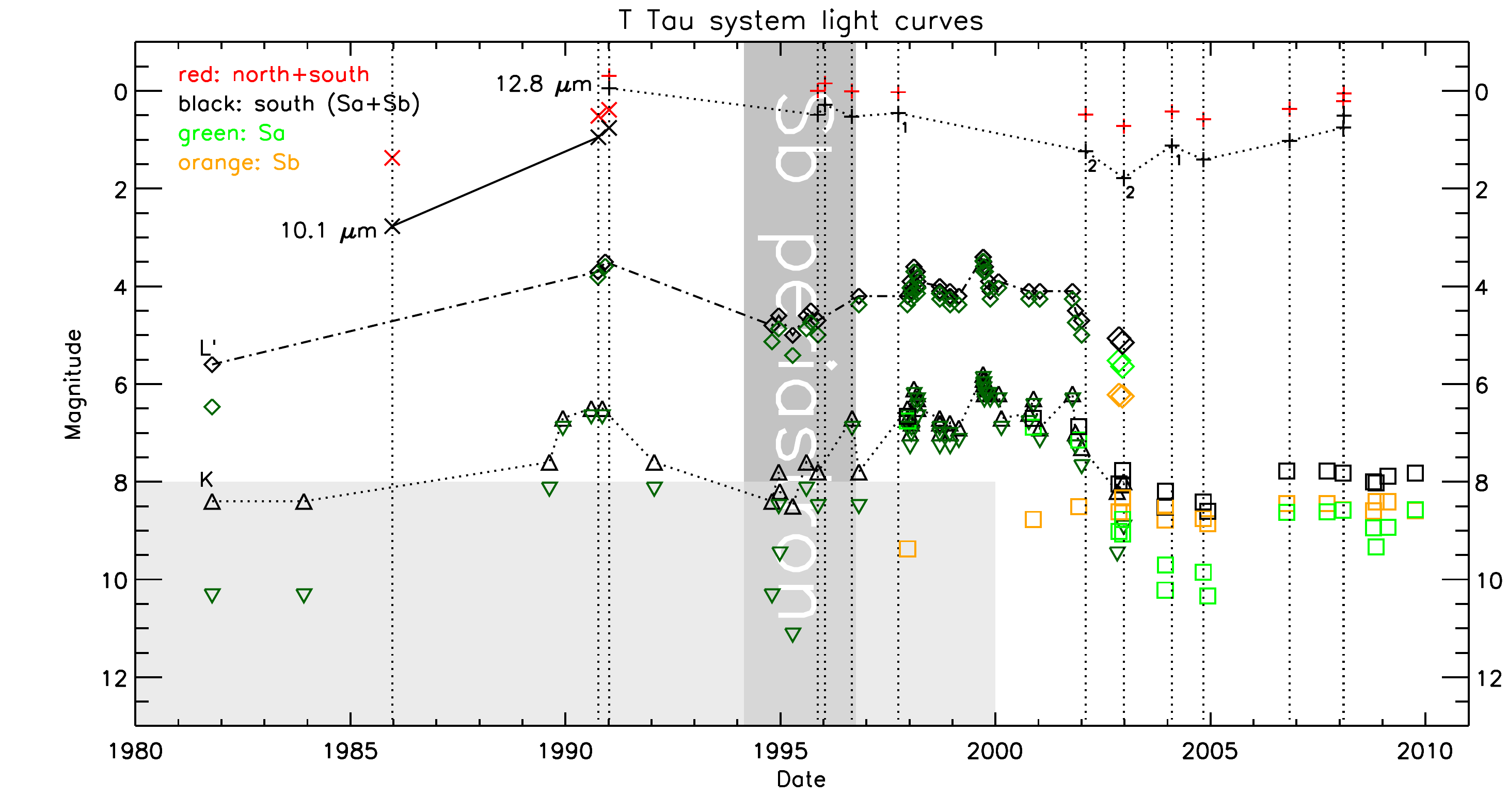}
\caption{\label{fig:light_curve} Updated light curves of various components of the T\,Tau system at various infrared wavelengths. References for all measurements shown are given in \tab\,\ref{tab:IR_photometry}.
\emph{At the top} (roughly between 0 and 3 magnitudes) are N-band measurements. Two epochs in which the observations were space based and did not spatially resolve the N-S pair are indicated with '1' sub-scripts. The two epochs of TIMMI\,2 observations that only very marginally resolve the N-S pair are marked with '2' sub-scripts. In red we plot the total system brightness (N+S), in black we plot the magnitude of the T\,Tau\,S (sum of Sa and Sb). 
\emph{In the middle} of the graph (roughly between 3 and 6 magnitudes) we show \Lp \ magnitudes of T\,Tau\,S. The black diamonds show the total flux of T\,Tau\,S, the green diamonds show the flux of Sa only. The observations of the last two epochs (large diamonds) spatially resolved Sa and Sb \citep{2005ApJ...628..832D,2007AJ....134..359H}, here the magnitude of Sb is indicated in orange. 
\emph{At the bottom} of the graph (roughly at 6 magnitudes and below) we show K-band magnitudes of T\,Tau\,S. Triangles indicate measurements in which the Sa-Sb pair was not spatially resolved, green and orange squares indicate the magnitudes of Sa and Sb, respectively, deduced from spatially resolved observations (Sa+Sb in black).}
\end{figure*}

Figure\,\ref{fig:light_curve} shows a compilation of photometric measurements of the T\,Tau system between 2 and 13~\mum, spanning a time interval of nearly three decades. The data shown consist mostly of available literature photometry, supplemented by the new measurements presented in this work. Similar curves showing a subset of these data have been presented earlier \citep{1991AJ....102.2066G,2004ApJ...614..235B}. The references for all data shown are given in \tab\,\ref{tab:IR_photometry}.

For the N-band fluxes (10.1\,\mum \ broad band fluxes for the earliest epochs, continuum fluxes at or near 12.8\,\mum \ for the epochs since the early 1990s) we show both the total system flux (N+Sa+Sb) and the flux of T\,Tau\,S (Sa+Sb) only. The N-band fluxes were converted to magnitudes using zeropoints of 41.0\,Jy at 10.1\,\mum \ and 25.6\,Jy at 12.8\,\mum. Note that only the fluxes at 12.8\,\mum \ are used in the further analysis. For the \Lp \ magnitudes we separately show the total flux of T\,Tau\,S (Sa+Sb) and the brightness of Sa alone. To obtain the fluxes of Sa from measurements that did not spatially resolve the Sa-Sb pair \citep[][see \tab\,\ref{tab:IR_photometry}]{2004ApJ...614..235B} we assumed that Sb was constant at \Lp=6.25\,mag. In two epochs of AO-assisted imaging the \Lp \ fluxes of Sa and Sb could be determined directly \citep[][]{2005ApJ...628..832D,2007AJ....134..359H}. Also for the K-band we show the total flux of T\,Tau\,S (Sa+Sb) and the flux from Sa alone separately. Besides a large number of spatially unresolved measurements, more than a dozen AO-assisted observations are shown that spatially separate the Sa-Sb pair.

\subsection{The color-magnitude behavior of the long-term variability}
\label{sec:color_magnitude_behavior}
The multi-wavelength infrared light curve presented in \fig\,\ref{fig:light_curve} allows us to investigate the relation between infrared brightness and SED shape of the infrared companion. The first comprehensive study of this kind was presented by \cite{2004ApJ...614..235B}, who presented a K$-$\Lp \ vs. K color-magnitude diagram of T\,Tau\,S and showed that its near-infrared color varies with brightness in accordance to the ISM extinction law. In \fig\,\ref{fig:KminLp} we show a CMD similar to that of Beck et al., except that we plot the magnitudes of Sa only (the Beck et al. diagram shows T\,Tau\,S, i.e. the sum of Sa and Sb).

In \fig\,\ref{fig:KminN} we extend the study of \cite{2004ApJ...614..235B} to longer wavelengths by compiling a K-[12.8] vs. K color-magnitude diagram of Sa. The 2.2\,\mum \ and 12.8\,\mum \ data were generally not obtained simultaneously and have time lags of up to nearly two months. Hence this discussion is necessarily limited to the long-term variability. Because Sa has been observed to vary substantially on timescales of days, we assigned uncertainties to the data points to account for the short-term variability in an approximate sense as follows. We took the fastest observed variations at 12.8\,\mum \ (0.26\,mag in four days, this work) and scaled this with the square root of the time lag between the 2.2 and the 12.8\,\mum \ observations. This choice of scaling is somewhat arbitrary but reflects the apparent randomness of the observed short-term fluctuations. 

\section{Discussion}
\label{sec:discussion}
In this section we will first discuss why the fast mid-infrared variability that we detected excludes variable extinction as a viable mechanism for the observed short-term mid-IR brightness fluctuations in T\,Tau\,S, {and argue that these must instead be due to variations in intrinsic luminosity.} Then we will present a radiative transfer disk model of Sa, including both an active accretion and a passive reprocessing component, that qualitatively reproduces the observed long-term color-magnitude variations when the accretion rate is varied. Thus we will show that variable accretion may also explain the long-term variability, but do not exclude variable extinction as the prime mechanism responsible for the long-term variations. Lastly, we sketch a tentative scenario for T\,Tau\,S, in which the \simil15\,yr period of irregular and enhanced brightness that we witnessed in the recent past is induced by the periastron passage of Sb, gravitationally disturbing the Sa disk and triggering an accretion outburst that may be reminiscent of EXOR variables.

\subsection{Can the observed short-term brightness fluctuations be caused by variable extinction?}
\label{sec:variable_extinction}

For variable extinction to be a viable cause of the observed brightness fluctuations, it is a necessary condition that the opaque medium causing the extinction can (un-) cover the emitting region \emph{within the timescale of the variations}. Because only Sa shows large brightness variations whereas T\,Tau\,N and Sb do not, one may assume the absorbing "screen" to be local to the system, i.e. at a distance of \simil148\,pc. Thus we can estimate the \emph{minimum speed} at which the screen must travel to cause the observed variations. To this purpose, we will first calculate the minimum size of the emitting region.

Our sole constraint is that T\,Tau\,S brightened by +3.3\,Jy within 94.5~hours at 12.8\,\mum. Because at 12.8\,\mum \ the emission is completely dominated by thermal emission from dust grains, the maximum intensity the emitting region can have is that of a \simil1500\,K black body (at higher temperatures, the dust evaporates and the material loses its IR opacity). Thus the minimum solid angle of the region that needs to be (un-) covered is:
 $$ 
 \Omega = F_{\nu}/B_{\nu}(1500\,\rm{K}) \approx 1.9\times10^{-16} \ {\rm sr},
 $$
 where $F_{\nu}$\,=\,3.3\,Jy, and $B_{\nu}$ denotes the Planck function (evaluated at 12.8\,\mum). At the system distance of 148\,pc, and approximating the emitting region by a circle\footnote{The argument is qualitatively independent of the exact shape of the emitting region and we adopt the simplest possible geometry.}, this yields a diameter of \simil0.48\,AU. The minimum required speed for the absorbing screen to (un-) cover this region within the available four days then becomes \simil210\,km\,s$^{-1}$.
 
This velocity is much higher than the velocities one may expect in the close environment of T\,Tau\,Sa. The Kepler speed at 0.24\,AU, the absolute minimum distance from the star at which the screen may be positioned in order to cover a region of 0.48\,AU in diameter, is \simil90\,km\,s$^{-1}$, assuming a mass of 2.2\,\Msun \ for Sa \citep{2008A&A...482..929K,2008JPhCS.131a2028K}. This velocity corresponds to a circular orbit, screens orbiting on eccentric orbits may reach velocities of up to \simil130\,km\,s$^{-1}$ at this distance. If the absorbing screen would be on an eccentric orbit, though, it would itself become warmer and brighter at 12.8\,\mum \ on closest approach, thus at least partially compensating for its own dimming effect. Thus, extinction caused by a "structure" existing within the disk, e.g. a warp or spiral arm, cannot be responsible for the observed IR variability. Moreover, the velocity dispersion within a molecular cloud is only a few km\,s$^{-1}$ \citep[e.g.][]{1981MNRAS.194..809L}, which excludes dust clouds in relatively close vicinity of the star, but not directly related to it, as viable absorbing screens.

In the above analysis we have made several simplifications, which were all chosen to \emph{lower} the required velocities for the observing screen, i.e. to favor the variable extinction scenario. The goal was to show that variable extinction can be ruled out, even with these unrealistically conservative assumptions:

\begin{itemize}
\item[$\bullet$] The required velocity of the obscuring screen that we derived is the absolute minimum value that does not violate fundamental laws of physics (Planck's or Kepler's law), under the assumption that the observed IR continuum radiation is thermal dust emission. However, in reality, the region of the disk that emits the bulk of the flux at 12.8\,\mum \ is likely much larger, leading to a correspondingly higher minimum required velocity. For a typical disk model around an object with approximately the stellar parameters of T\,Tau\,Sa, the "size" of the disk (here defined as the region within which 75\% of the flux at 12.8\,\mum \ is emitted) is about 15\,AU in radius \citep[e.g.][]{2005A&A...441..563V}. In the special case of T\,Tau\,Sa, where the disk is likely truncated on the outside at a radius of 3$-$5\,AU due to tidal interactions with T\,Tau\,Sb, the region we see at 12.8\,$\mu$m probably encompasses the entire disk. Conservatively assuming an outer radius of 3\,AU for the disk, an absorbing screen would need to travel at \simil700\,km\,s$^{-1}$ in order to (un-) cover 26\% of the emitting region within four days, whereas the Kepler velocity at this radius is only \simil26\,km\,s$^{-1}$.

\item[$\bullet$] We assumed the absorbing screen to be 100\% opaque at 12.8\,\mum. If the screen's transparency at this wavelength is considerable, a larger area needs to be (un-) covered within the available time and the required velocity becomes correspondingly higher.
\end{itemize}
We conclude that variable extinction \emph{cannot} cause the short-term brightness fluctuations observed at 12.8\,\mum. The absolute minimum velocity required for an absorbing screen to (un-) cover the emitting region within the available time is about a factor of 2 higher than the highest possible speed such a screen could have at the relevant location. A more realistic estimate yields a discrepancy of a factor of \simil20-30 between the required and the physically possible velocity.

Note that if we apply the same simple model to the fastest variations detected by \cite{2004ApJ...614..235B} \emph{at 2.2\,\mum} \ (\simil0.9\,Jy in seven days)\footnote{The brightness of T\,Tau\,S increases from K=7.4\,mag on 1997 December 6 to K=6.5\,mag seven days later \citep{2004ApJ...614..235B}. Assuming Sb was constant at K=8.6\,mag, this implies that Sa went from K=7.84\,mag to K=6.67\,mag, i.e. it brightened by \simil1.2\,mag, or \simil0.9\,Jy.}, we cannot exclude variable extinction: a 1500\,K blackbody emitting the corresponding flux would have a diameter of \simil0.15\,AU and the minimum speed needed for an opaque screen to (un-) cover this region within 7 days is only \simil37\,\kms. This is substantially below the corresponding Kepler speed of \simil160\,\kms, and thus variable extinction is a perfectly viable explanation for the observed variations. Considering that the stellar photosphere may contribute significantly to the total flux at 2.2\,\mum, the case for variable extinction becomes even stronger. \emph{Only the long wavelength of our VISIR observations allows rejection of the variable extinction hypothesis}.

\begin{figure}[t]
\hspace{-0.7cm}
\includegraphics[width=10.0cm,angle=0]{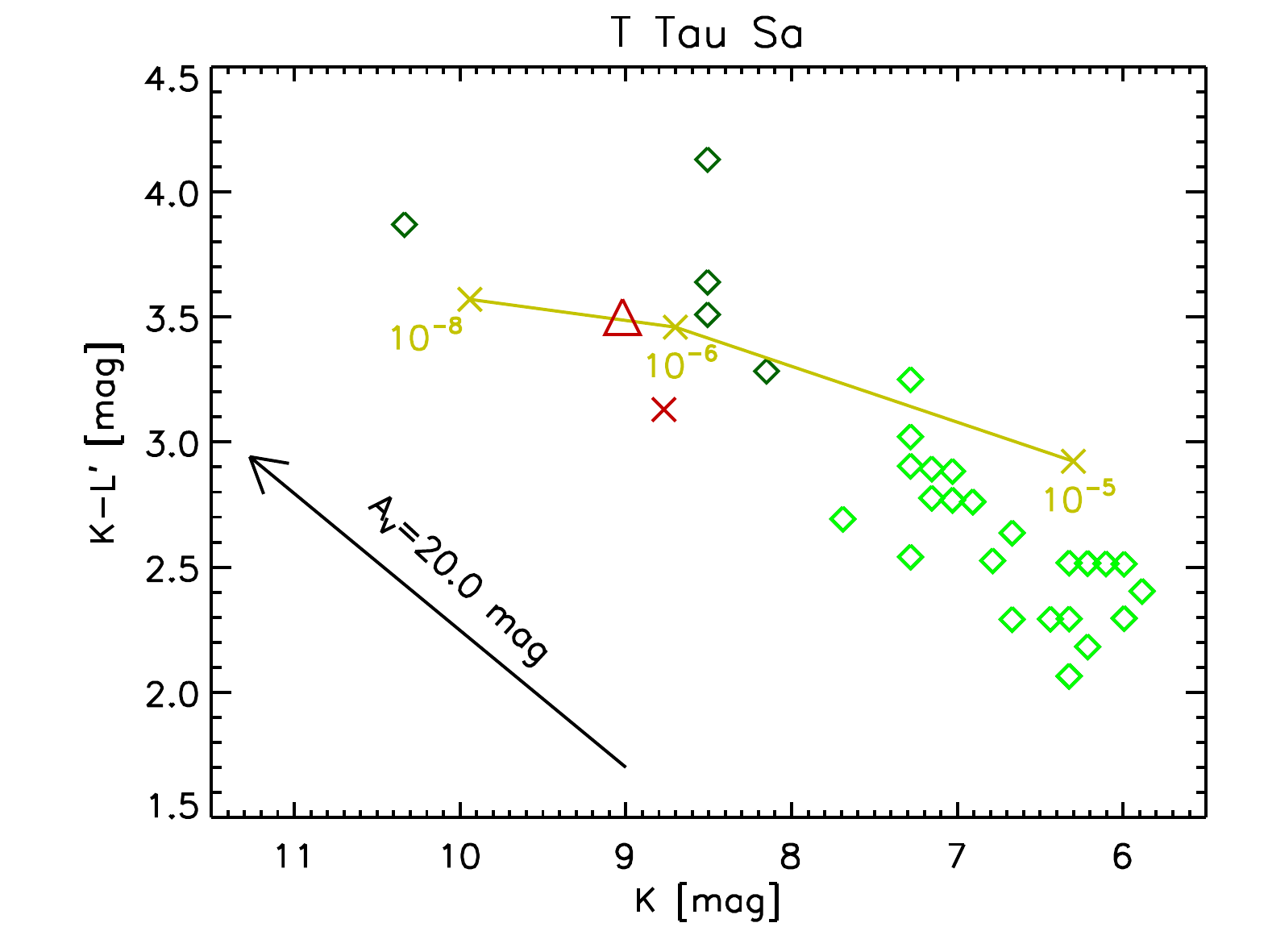}
\caption{\label{fig:KminLp} K-\Lp \ vs. K color-magnitude diagram similar to that presented by \cite{2004ApJ...614..235B}, except that we plot only Sa (\cite{2004ApJ...614..235B} plot the sum of Sa and Sb). The diamonds represent \emph{indirect measurements} calculated from the data by \cite{2004ApJ...614..235B}, in which Sa-Sb remained spatially unresolved, assuming that Sb was of constant brightness at K=8.6\,mag and \Lp=6.25\,mag, respectively. The diamonds in the upper left part of the diagram, in a darker shade, are less reliable because Sa was comparatively faint during the epochs of these measurements. The cross and triangle are from \cite{2005ApJ...628..832D} and \cite{2007AJ....134..359H}, respectively, whose measurements \emph{did} spatially resolve the Sa-Sb pair. Over-plotted are radiative transfer model magnitudes, labeled with their respective accretion rates in \msunyr, see \sek\,\ref{sec:radiativeTransfer} and \tab\,\ref{tab:model_parameters}.}
\end{figure}

\begin{figure}[t]
\hspace{-0.7cm}
\includegraphics[width=10.0cm,angle=0]{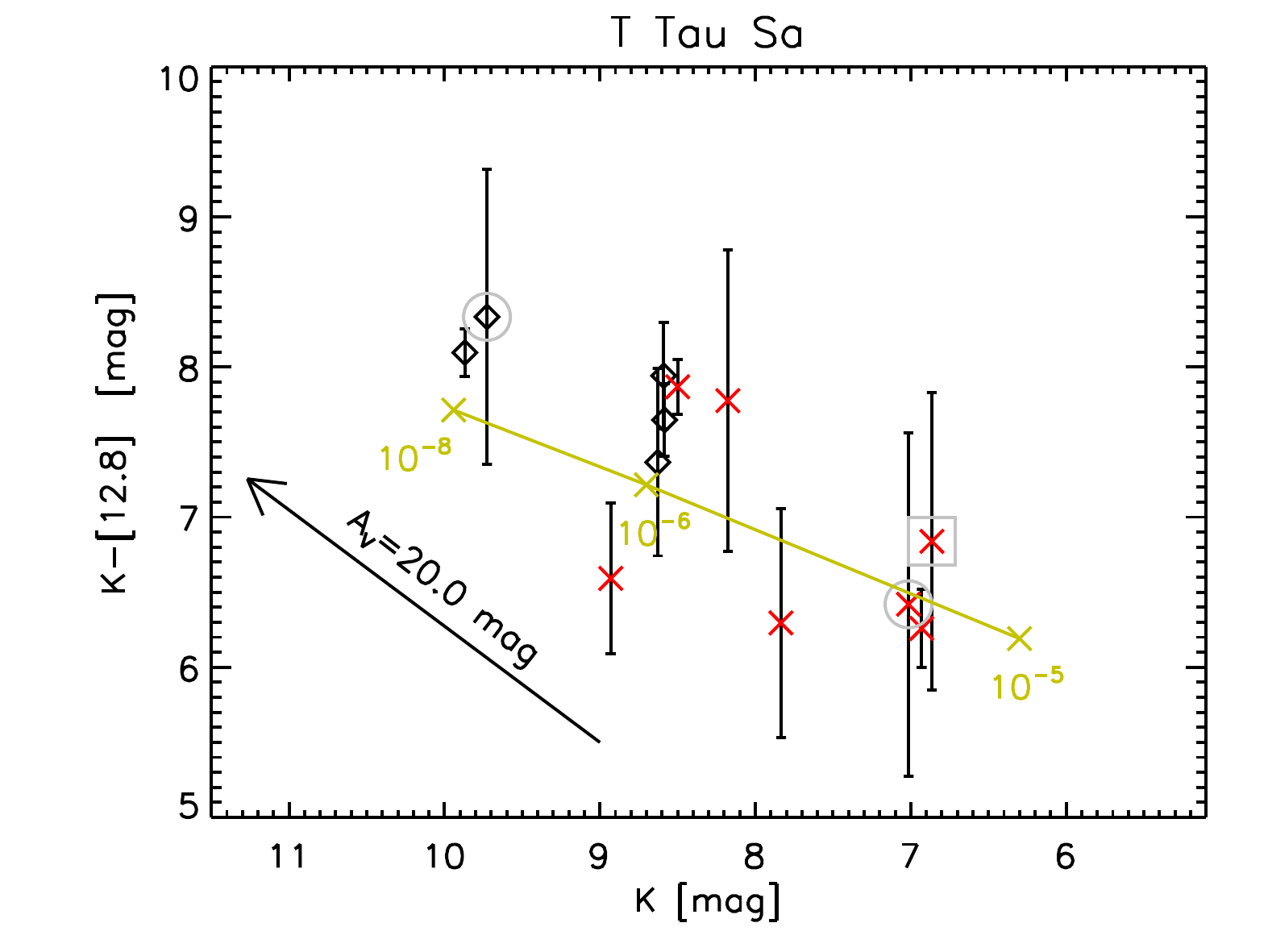}
\caption{\label{fig:KminN} K-12.8 vs. K color-magnitude diagram of T\,Tau\,Sa. 
 The diamonds represent measurements in which the Sa-Sb pair was spatially resolved in the K-band, crosses denote measurements in which we calculate the K-band magnitude of Sa from the total flux of T\,Tau\,S assuming Sb has K=8.6. The N-S pair was spatially resolved at 12.8\,\mum, except for the two points encircled in grey, for which the 12.8\,\mum \ flux was calculated from the total flux of T\,Tau assuming T\,Tau\,N has a flux of 8.2\,Jy. The point surrounded by a grey box represents a broad band measurement centered at 12.5\,\mum \ by \cite{1991AJ....102.2066G}. The uncertainties arise mostly because the near- and mid-infrared observations were not performed simultaneously (see \sek~\ref{sec:color_magnitude_behavior}). Over-plotted are radiative transfer model magnitudes, labeled with their respective accretion rates in \msunyr, see \sek\,\ref{sec:radiativeTransfer} and \tab\,\ref{tab:model_parameters}.}
\end{figure}

\begin{table}[b]
\caption{\label{tab:model_parameters} Summary of the parameters of the radiative transfer model described in \sek\,\ref{sec:radiativeTransfer} and \figs~\ref{fig:KminLp}, \ref{fig:KminN}, and \ref{fig:Sa_SEDs}.}

\begin{tabular}{lccc}

parameter      &  symbol        &  value                           & unit    \\
\hline

stellar radius         & $R_*$  &  3.5  &  \Rsun \\
stellar mass           & $M_*$  &  2.2  &  \Msun \\
effective temperature  & $T_{\rm{eff}}$  &  5500 &  K  \\
stellar luminosity     & $L_*$  &  10   & \Lsun \\

\vspace{-0.15cm} \\

disk mass      & $M_{\rm{disk}}$ &  10$^{-3}$ & \Msun \\
exponent surface density  & $p$          &  -1  &   \\
inner radius passive disk & $R_{\rm{in}}$ &  0.5 & AU \\
outer radius passive disk & $R_{\rm{out}}$ &  5.0 & AU \\
scale height at inner edge & $H_0$    &  0.021 & AU \\ 
flaring index              & $\gamma$ &  2/7 &    \\

\vspace{-0.15cm} \\

accretion rate &       \Mdotacc &  10$^{-8}$, 10$^{-6}$, 10$^{-5}$ & \msunyr \\
inner radius active disk  &  $R_{\rm{in, SS}}$ &  0.036 & AU \\  
outer radius variable \Mdotacc  &  $R_{\rm{out, acc}}$ &  0.5 & AU \\

\vspace{-0.15cm} \\

inclination               & $i$       &  74     &  deg \\
foreground extinction     & $A_{\rm{V}}$ & 15      &  mag \\

\hline
\end{tabular}

\end{table}

\subsection{Can the observed long-term brightness fluctuations be caused by variable accretion?}
\label{sec:radiativeTransfer}

In the previous section we showed that apparent brightness fluctuations due to variable extinction cannot be the cause of the observed short-term mid-IR photometric variations of T\,Tau\,Sa. Therefore the variability must be due to intrinsic changes in the luminosity of Sa. A high and strongly variable accretion rate would naturally lead to such luminosity changes in a young star like Sa, as already argued by \cite{1991AJ....102.2066G}. Thus, variable accretion is the prime candidate mechanism to explain the observed short-term variability. One may ask whether it can explain the observed long-term photometric variations, which are of much larger amplitude, as well. The variable accretion mechanism has previously been deemed unfit to explain the ``bluer when brighter'' color-magnitude behavior of the long-term brightness fluctuations \citep{2004ApJ...614..235B}. In this section we will show that constructing a disk model that qualitatively reproduces the observed color-magnitude behavior \emph{is} in fact possible, with parameters that remain in a realistic range.

Our model of the Sa disk consists of two components: an ``active'' inner disk in which luminosity is generated by the release of gravitational energy from accreting material, and a ``passive'' outer disk that is irradiated by the central star and the inner disk, and reprocesses the absorbed optical/NIR photons into longer wavelength infrared radiation. The model parameters are briefly discussed here and are also listed in \tab~\ref{tab:model_parameters}. They are chosen so that they represent Sa and its disk to the best of current knowledge. In particular the small outer radius of the disk and the near edge-on orientation are key features \citep[e.g.][]{2005ApJ...628..832D}.

The active, inner part is modeled with an optically thick, geometrically thin accretion disk with a radial temperature profile following the \cite{1973A&A....24..337S} model. In addition, a ``hot-spot'' covering 1\% of the star, representing the accretion shock of infalling material hitting the stellar surface, is included by adding a blackbody of the corresponding temperature and luminosity to the stellar spectrum. For the passive, outer disk we used the 2D radiative transfer code {\sc radmc} \citep[][]{2004A&A...417..159D} to calculate the temperature structure and emergent infrared SED. The passive disk part has an inner radius of $R_{\rm{in}}$\,$=$\,0.5\,AU, an outer radius of $R_{\rm{out}}$\,$=$\,5\,AU, a mass of $M_{\rm{disk}}$\,$=$\,$10^{-3}$\,\Msun \ (assuming a gas to dust ratio of 100) with a radial surface density profile of $\Sigma$\,$\propto$\,$R^{-1}$, a flared geometry with a flaring index\footnote{The flaring index $\gamma$ is defined so that the ratio of the disk scale height $H$ and the distance to the central star $R$ scales like $H/R$\,$\propto$\,$R^{\gamma}$.} of $\gamma$\,$=$\,2/7 and a scale height of $H_0$\,$=$\,0.021\,AU at the inner disk edge of the passive, re-processing disk ($H/R$\,$=$\,0.042, chosen to be the value corresponding to hydrostatic equilibrium in the "high" state). The outer radius is much smaller than that of typical circumstellar disks because of tidal truncation by Sb, whose highly eccentric orbit brings it to within \simil12\,AU from Sa at closest approach. This naturally explains the non-detection of the Sa disk at mm~wavelengths by \cite{1997ApJ...490L..99H} and \cite{1998ApJ...505..358A}, as argued before by \cite{2000ApJ...531L.147K}. 

By varying the accretion rate, we change the amount of energy that is released in the center of the system, heating the circumstellar material. The luminosity of the stellar photosphere dominates at low accretion rates, providing a "base level" for the energy output. At accretion rates above \simil7$\times$10$^{-7}$\,\msunyr \ accretion becomes the main energy source and the heating rate of the circumstellar material scales roughly linearly with the accretion rate. Variations in the central luminosity lead to essentially instantaneous variations in the temperature of the circumstellar material and to corresponding changes in the infrared emission because the dust grains are small and attain their new equilibrium temperature practically instantly if the radiation field changes, and the light travel time to the outer disk edge is only about 30 minutes.

For the luminosity of the central star itself, we adopt a value of 10\,\Lsun. This is lower by a factor of a few compared to the estimate by \cite{2005ApJ...628..832D}, who base their value on a mass estimate of 2.5-3\,\Msun \ for Sa. However, dynamical modeling of the Sa-Sb orbit including more recent data yields a mass estimate of $\sim$2.2\,\Msun \ for Sa \citep{2008A&A...482..929K,2008JPhCS.131a2028K}, which is essentially identical to the mass of T\,Tau\,N \citep[2.1\,\Msun,][]{2001ApJ...556..265W}. The bolometric luminosity of T\,Tau\,N is relatively well determined to be 7.3$^{+1.3}_{-1.1}$\,\Lsun \ \citep{2001ApJ...556..265W}, and assuming the stars are co-eval therefore yields a luminosity of $\lesssim$10\,\Lsun \ for Sa. A lower stellar luminosity is favorable for a scenario in which the IR variability is explained by variable accretion: the required extra luminosity in ``high states'' is always relative to the stellar photosphere and thus directly scales with the stellar luminosity. The stellar parameters that we chose correspond to a 2\,Myr-old star of 2.2\,\Msun \ according to the pre-main sequence evolutionary tracks by \cite{2008ApJS..178...89D}.

\begin{figure}[t]
\hspace{-0.7cm}
\includegraphics[width=10.0cm,angle=0]{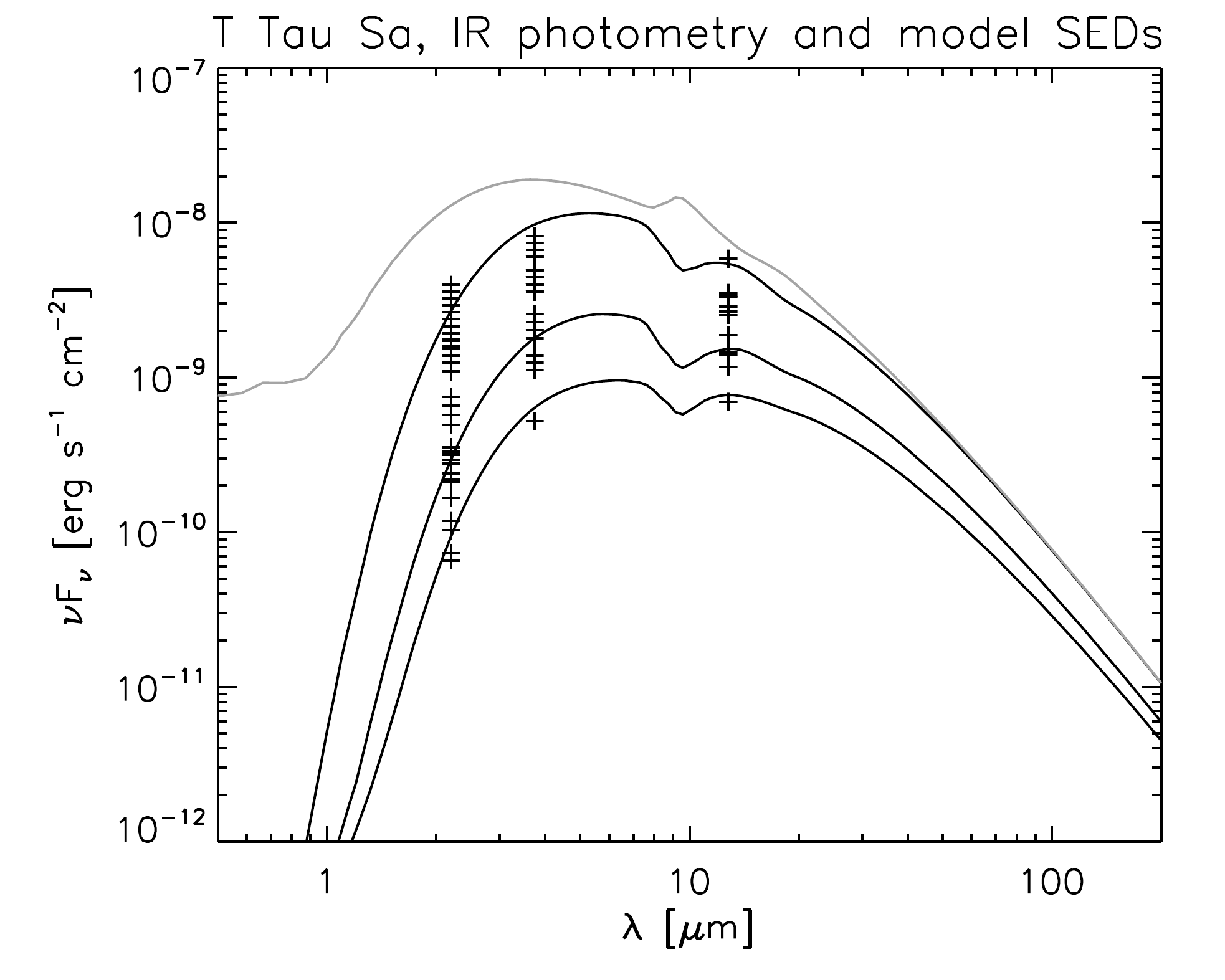}
\caption{\label{fig:Sa_SEDs} Observed infrared photometry of Sa, with radiative transfer model SEDs over-plotted. The photometry is the same as in \figs~\ref{fig:KminLp} and \ref{fig:KminN}. The model SED curves are for accretion rates of 10$^{-5}$, 10$^{-6}$, and 10$^{-8}$\,\msunyr \ (top to bottom black curve), with the other model parameters as given in \sek~\ref{sec:radiativeTransfer} and \tab~\ref{tab:model_parameters}. Also shown, with a grey curve, is the 10$^{-5}$\,\msunyr-model without foreground extinction.}
\end{figure}

We performed model calculations for mass accretion rates of 10$^{-8}$, 10$^{-6}$, and 10$^{-5}$~\msunyr. For the lowest accretion rate, the accretion luminosity amounts to $\lesssim$2\% of the stellar luminosity and the resulting fluxes are essentially identical to those obtained with a zero accretion rate. 
We calculated models for a range of nearly edge-on inclinations. Because our objective is to demonstrate that a variable accretion model is in principle capable of reproducing the observed color-magnitude behavior, we searched for a model that gives a fair match to the observations but did not comprehensively explore parameter space. A model with parameters as listed earlier in this section and in \tab~\ref{tab:model_parameters}, seen at an inclination of 74\degree \ and with a foreground extinction of 15\,mag agrees fairly well with the observations (the extinction matches the value of the foreground extinction observed toward Sb \citep{2005ApJ...628..832D}). This particular inclination yields the best fit for the set of model parameters that we chose, but other choices of disk parameters will result in slightly different best-fit inclinations. Thus, our model does not accurately constrain the inclination of the Sa disk, but we stress that a near edge-on inclination is required to obtain the correct color-magnitude behavior of the variations, regardless of the specific choice of disk parameters.

In \fig\,\ref{fig:Sa_SEDs} we show the model SEDs for accretion rates of 10$^{-8}$, 10$^{-6}$, and 10$^{-5}$~\msunyr, as well as the observed infrared photometry. We also show the model for the highest accretion rate as it would appear if the foreground extinction is omitted (grey curve). In \figs~\ref{fig:KminLp} and \ref{fig:KminN} we showed the observed color-magnitude behavior of the photometric variations of Sa. The model magnitudes are over-plotted and labeled with their respective accretion rates. The results of our modeling efforts can be summarized as follows:
    
\begin{itemize}
\item[$\bullet$] the near-infrared color and brightness variations are qualitatively reproduced: we obtain a "redder when fainter" behavior in a K-\Lp \ vs. K color-magnitude plot (see \fig\,\ref{fig:KminLp}). The slope of the relation in the model is somewhat shallower (i.e. closer to grey) than observed.
\item[$\bullet$] the observed K-[12.8] \ vs. K color-magnitude behavior is qualitatively reproduced. The slope of the modeled relation fits the observations well. The considerable observed scatter can at least in part be attributed to the time lag between the near- and mid-infrared observations (see \fig\,\ref{fig:KminN}).
\item[$\bullet$] the disk model always produces a 10\,\mum \ silicate in emission, even at 90\degree \ (edge-on) inclination. Additional foreground extinction is needed to explain the observed silicate absorption feature.
\end{itemize}

Why does our model reproduce the observed ``bluer when brighter'' behavior, whereas the previous model considered in the context of T\,Tau\,S \citep[][]{1997ApJ...481..912C,2004ApJ...614..235B} did not? In the Calvet model the central star is not highly extincted. At low accretion rates, the very blue stellar photospheric emission dominates the near-infrared spectrum. As the accretion rate is increased, the comparatively red disk spectrum becomes more prominent and dominates the near-infrared emission above 10$^{-7}$\,\msunyr. The changing star-disk contrast causes a ``redder when brighter'' color-magnitude behavior with varying mass accretion rate. In contrast, in our highly reddened model of the Sa star and disk system, the photosphere never substantially contributes directly to the near-infrared flux, because it is obscured by the Sa disk. At low accretion rates, the emission is dominated by the far side of the inner edge of the passive disk part, irradiated by the central star. With increasing accretion rate, the inner edge of the passive disk becomes hotter and, in addition, the comparatively blue active Shakura-Sunyaev disk contributes stronger to the near-infrared spectrum. The combined effect is a ``bluer when brighter'' color-magnitude behavior.

The variable accretion mechanism has been proposed previously, based on the apparent grey behavior of the observed near- to mid-infrared flux variations \citep{1991AJ....102.2066G,2002ApJ...568..771D}. However, based on a much larger data set we show that the large, long-term flux variations at 2.2 and 12.8\,\mum \ are accompanied by substantial color variations (\fig\,\ref{fig:KminN}), which can be reproduced by our variable accretion model. However, these data can also be explained with variable foreground extinction (see the reddening vectors drawn in \figs~\ref{fig:KminLp} and \ref{fig:KminN}). The total amplitude of the foreground extinction required to cover the observed brightness range is $\Delta$\Av\simil40\,mag.

\subsubsection{Variable extinction causing variable disk-illumination?}
In the previous paragraphs we showed that variable extinction in its simplest form, i.e. absorbing material passing in front of a source of approximately constant intrinsic luminosity, cannot cause the observed fast mid-IR brightness variations. However, the color-magnitude behavior of the large, multi-magnitude variations that have been observed during the last two decades, \emph{can} be explained with variable extinction (see \sek~\ref{sec:radiativeTransfer}). The variable accretion model is in principle capable of explaining both the detected sub-magnitude fast variations, as well as the presumably slower multi-magnitude variations. Here we will consider a conceptually different realization of the variable extinction scheme, in which we attempt to explain both the short- and long-term variations with extinction alone.

To this purpose, we will introduce ``central extinction events'': variable absorption columns close to the central star that cause substantial changes in the illumination of the disk surface at larger radii with optical and near-infrared radiation. This could in principle lead to fast variations at mid-IR wavelengths because the material causing the variable disk illumination resides much closer to the central star than the material emitting the bulk of the mid-IR radiation, thereby circumventing the dynamical timescale problem encountered in the ``classical'' variable extinction scenario.

In order to explore this scheme in a more quantitative sense, we constructed a passive disk model for Sa, in which we introduce variable absorption columns by increasing the scale height at the inner disk edge. The disk parameters\footnote{$R_{\rm{in}}$\,$=$\,0.5\,AU, $R_{\rm{out}}$\,$=$\,5\,AU, $M_{\rm{disk}}$\,$=$\,$10^{-3}$\,\Msun, $\Sigma$\,$\propto$\,$R^{-1}$, $\gamma$\,$=$\,2/7, $H_0$\,$=$\,0.021\,AU at 0.5\,AU.} are the same as those of the model presented in \sek\,\ref{sec:radiativeTransfer}, but as the central illuminating source we replaced the 10\,\Lsun \ star plus variable accretion component with a star-only configuration with a constant luminosity of 40\,\Lsun \ \citep{2005ApJ...628..832D}.

We considered both a scenario in which the scale height at the inner disk edge is time variable, and one in which a fixed ``structure'' rotates at the inner disk edge with the Keplerian speed, casting a shadow over the outer disk that rotates along with the same angular velocity. For the former experiment, we artificially increased the scale height of the disk between $R_{\rm{in}}$ and 2$R_{\rm{in}}$, at all azimuthal angles. If we increase the scale height by 55\%, with respect to the disk parameters given in the previous paragraph, the flux at 12.8\,\mum \ decreases by 0.26\,mag because the inner disk somewhat shadows the regions at larger radii. The \simil0.26\,mag brightening at 12.8\,\mum \ that we observed in February 2008 may conversely be explained by the opposite change in inner rim structure: a reduction of the pressure scale height from \simil0.034\,AU to \simil0.022\,AU within $\lesssim$4\,days. The height above the midplane where the disk becomes optically thin to the stellar irradiation\footnote{Here this is taken to be the height where the optical depth in the radial direction at $\lambda$\,$=$\,0.55\,\mum \ falls below unity.} then decreases from \simil0.16\,AU to \simil0.11\,AU, i.e. by more than twice the pressure scale height corresponding to hydrostatic equilibrium. The orbital period at 0.5\,AU from a 2.2\,\Msun \ star is \simil87~days, and thus vertical movements on the order of one scale height require roughly 20 days, which is much longer than our observational constraint. Thus we conclude that changes in the inner disk structure cannot cause the observed fast mid-IR brightness variations.

A somewhat different approach is to take a fixed enhancement of the disk scale-height at the inner rim over only part of the azimuthal range, causing a ``shadow'' to be cast over the disk behind it. This shadow will move over the outer disk with an angular velocity corresponding to the Kepler speed at the inner edge, and thus cause a ``cool region'' at larger radii that moves at locally vastly super-keplerian velocities over the disk surface. For near edge-on inclinations, where the mid-IR flux we see is dominated by the far side of the disk, this can cause substantial dimming. We used the 3D version of {\sc radmc} (Dullemond et al. in prep.) to calculate light curves of a system in which the scale height at the inner rim is increased with respect to hydrostatic equilibrium over a limited azimuthal range\footnote{Over $\pi$ radians in azimuth the scale height is increased with a sinusoidal shape, so that at the central azimuthal angle the scale height is twice the hydrostatic value, and at angles of $-\pi$/2 and $+\pi$/2 with respect to the maximum it reaches the hydrostatic value.}. With this we can achieve brightness variations of $\lesssim$0.15\,mag on timescales of \simil20~days. Because the enhancement of the scale height is unrealistically large and yet the brightness variations are substantially smaller and slower than observed, we do not consider this scenario to be a viable explanation for our observations.

\vspace{0.15cm}

We conclude that variable accretion is the prime mechanism responsible for the observed \emph{short-term} mid-IR variability, and exclude a significant role for variable extinction in this matter, as explained in \sek~\ref{sec:variable_extinction}. The short-term near-IR photometric variations are naturally also explained by variable accretion, though the existing data do not strictly exclude variable extinction as their cause. Based on the hitherto available data, both variable accretion and variable extinction are viable mechanisms to explain the \emph{long-term} photometric variations and the associated color-magnitude behavior. Data presented in past studies, in particular a time-variable optical depth in the 3.05\,\mum \ ice absorption feature \citep[][]{2001ApJ...551.1031B}, do suggest that at least part of the variability can be attributed to variable extinction.

\subsubsection{The 10\,\mum \ silicate feature in Sa}

The fact that T\,Tau\,S shows a deep silicate absorption feature in the 10\,\mum \ spectral region \citep{1991AJ....102.2066G}, which was later shown to arise in Sa only \citep{2008ApJ...676.1082S,2009A&A...502..623R}, has been regarded as evidence that the 10\,\mum \ emission is dominated by a disk that is seen nearly edge-on from our vantage point \citep[e.g.][]{2005ApJ...628..832D}. Our radiative transfer model of the Sa disk, however, always shows a silicate feature in emission, even for an edge-on inclination (see the unreddened model curve in \fig\,\ref{fig:Sa_SEDs}, grey line). This is due to the "artificial" truncation of the Sa disk by the tidal influence of Sb, with an outer radius of 5\,AU in our model. In "normal " T\,Tauri disks, which are much larger, cold material in the outer disk regions obscures the warm inner disk when viewed at near edge-on inclinations, causing a silicate feature in absorption. In the special case of Sa, the entire disk is warm enough to emit in the 10\,\mum \ spectral region. The radiation that we receive is dominated by the optically thin upper disk layers, yielding a silicate feature in emission, regardless of the viewing angle. \cite{2009A&A...502..623R} reach the same conclusion based on their  radiative transfer model for Sa using the MC3D code. 
 Note that the shape and peak to continuum ratio of the silicate band depend on the grain size used. The small, sub-micron sized grains used in our model cause a relatively strong emission band peaking just short-ward of 10\,\mum. The use of larger grains results in a weaker, broader silicate feature, and one can make the silicate band essentially disappear from the spectrum by using very large ($\gtrsim$10\,\mum) grains. However, one never gets a silicate feature in absorption. Thus, an additional, external absorption component needs to be invoked. This is consistent with the fact that Sb, whose disk is thought to be seen more or less face-on \citep[e.g.][]{2005ApJ...628..832D}, suffers an extinction of \Av\simil15\,mag. The shape of the observed silicate absorption feature implies that the material causing the absorption consists predominantly of small ($\lesssim$1\,\mum) grains.

\subsection{A tentative scenario for T\,Tau\,Sa}
\label{sec:TTauS_scenario}
The light curves shown in \fig\,\ref{fig:light_curve} can qualitatively be described as follows\footnote{We will describe the K-band light curve because it has the best temporal sampling. The \Lp \ and 12.8\,\mum \ light curves follow the K-band (see also Figs.\,\ref{fig:KminLp} and \ref{fig:KminN}).}. From the beginning of the IR observations, in 1982, until 1989 T\,Tau\,S appears to have been in a ``low state''. Because there are only three measurements in this period, it is by no means certain that it has been faint during this whole period, but the available measurements all show T\,Tau\,S to be of essentially the same (low) brightness at 2.2\,\mum. Sometime in 1989, the brightness of the southern component started to increase strongly. Based on data from later epochs this increase, or at least the vast majority of it, is attributed to Sa only. During the period from \simil1991 to \simil1995 the temporal sampling was very poor, but the few available measurements suggest Sa to be relatively faint. From \simil1996 to late 2001 Sa has been in a ``high and variable'' state. This period has been very well sampled by \cite{2004ApJ...614..235B} and the K-band brightness varied between K\simil7.5\,mag and K\simil6\,mag. From late 2001 to late 2003 the brightness of Sa decreased to K\simil10.5\,mag. The temporal coverage of the observations is poor and thus we cannot assess how ``smooth'' this decrease was. Sometime between late 2004 and late 2006 the brightness of Sa rose again to K\simil9\,mag, and is currently varying around this value.

In view of the variable accretion scenario, the dramatic brightening events in the early and late 1990s and the interspersed episodes of irregular flux variations, of which we may be seeing the last twitch at the current time, could be interpreted as an ``outburst'' of accretion. In this respect, it is particularly interesting to compare the timing of the outburst with that of the periastron passage of the companion Sb. In the current best-fit solution of the Sa-Sb orbit, the closest approach occurred in June 1995, with an uncertainty of somewhat more than a year (1$\sigma$), as we have indicated in \fig\,\ref{fig:light_curve} with a grey vertical band. Thus, we have a period of enhanced activity around the time of closest approach of the companion, with the middle of the active period timed somewhat after periastron passage, and the strongest activity occurring in the second half of the active period. We will now argue that this behavior provides a good qualitative match to a scenario in which an accretion outburst in the center of the Sa disk is induced by the gravitational influence of Sb, tidally disturbing the Sa disk during periastron passage.

\begin{figure}[t]
\includegraphics[width=9.7cm,angle=0]{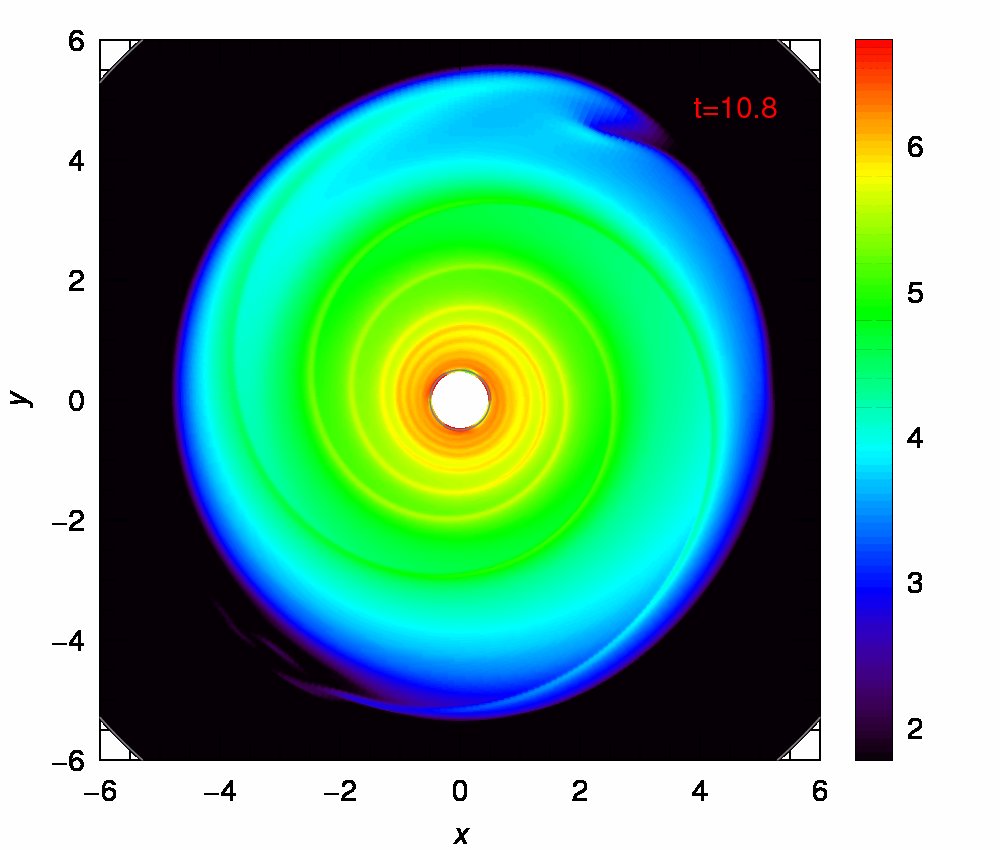}
\caption{\label{fig:kley_simulation} Two-dimensional temperature distribution of the Sa disk 10.8 yrs after the periastron passage of Sb, for a hydrodynamic simulation using the physical parameters of the T\,Tau\,S system. X and Y are spatial coordinates, and the disk is viewed face-on in this representation. To better bring out the entire range of temperatures we show $T[K]^{1/4}$, the colors cover the temperature range of 10 to 2000\,K.}
\end{figure}

When a young star and disk system has a low-mass companion, the latter will tidally truncate the disk around the primary to 
$\sim$1/3 times the semi-major axis of the orbit 
\citep[e.g.][]{1993ApJ...419..166A}. 
If the companion is on an eccentric orbit, its gravitational influence on the primary disk will be strongly time-dependent, being relatively weak during most of the orbit but culminating near periastron passage. \cite{2008A&A...486..617K} modeled the close binary system $\gamma$~Cephei in its youth, which is thought to have been very similar to the Sa-Sb system today. It has primary and secondary masses of \simil1.59\,\Msun \ and \simil0.38\,\Msun, respectively, a semi-major axis of \simil18\,AU, and an eccentricity of \simil0.36 \citep{2003ApJ...599.1383H}. In addition, the primary has a planetary companion with a mass of $M$\,sin\,$i$\,\simil\,1.7\,M$_{\rm{Jup}}$ \citep{2003ApJ...599.1383H}, which is, however, not of central importance to the current discussion. \cite{2008A&A...486..617K} performed hydro-dynamical simulations of the disk around the young $\gamma$\,Cep primary and considered the tidal effects of the low-mass secondary. The simulations reveal a nearly undisturbed disk at most orbital phases, when the companion is at relatively large distances. As the companion approaches the primary, however, two strong spiral shock waves are excited which propagate all the way to the inner disk (see \fig\,\ref{fig:kley_simulation}). This takes some time, and therefore the gravitational disturbance of the inner disk is not symmetrical in time with respect to the periastron passage, but lags somewhat. The strongest disturbances at the inner disk edge occur substantially after closest approach.

We speculate that such shock waves induce enhanced accretion in the innermost disk region of Sa when Sb is near periastron. For example, the mechanical energy dissipated in the shocks may cause a normally essentially neutral disk medium to be sufficiently ionized for MRI to become effective, naturally leading to strongly enhanced accretion. We have re-run the Kley \& Nelson model including viscous heating and radiative cooling for the parameters of the Sa-Sb system, and find the same behavior as seen in the $\gamma$\,Cep calculations\footnote{Note that the model discussed here and shown in \fig\,\ref{fig:kley_simulation} is a hydrodynamical calculation and is independent of the radiative transfer model discussed in \sek\,\ref{sec:radiativeTransfer}.}. The simulations provide a good qualitative match to the observations: a period of enhanced accretion activity near every periastron passage, whose peak has a substantial time lag with respect to the closest approach of the secondary. Quantitatively, the model shows much more modest variations in the accretion rate than inferred by the observations, and the time between periastron passage and maximum accretion activity is longer in the simulations by a factor of $\sim$3. The Kley \& Nelson model focuses on the dynamical evolution and mass loss of the disk at its outer edge, but does not attempt to treat the accretion physics at the inner disk edge in detail. A model with a more detailed treatment of the disk viscosity and accretion physics in the innermost disk regions is needed, which allows the spiral shock waves to propagate inward more quickly and can account for the required high accretion rates. This is beyond the scope of the current work, and whether our qualitative model can be refined to provide a quantitative match to the observations once all the relevant physics are included remains to be confirmed.

Here we pose the companion-induced accretion outburst as an idea to explain the observed behavior of T\,Tau\,S during the past decades, and provide a clear prediction that can be easily verified observationally: if this scenario is correct, Sa should return to its ``low state'' some time in the near future, and stay there until the next periastron passage of Sb. Exactly how long it will take for Sa to return to ``quiescence'' in detail depends on the actual accretion physics and inner disk parameters, neither of which we know. Thus, while it is technically trivial to test the proposed scenario with currently available observing facilities, it may in practice take a substantial number of years to convincingly falsify our scheme. Supportive evidence could be built up over the years if Sa is observed to remain faint, the final prove may come only many years from now if the next outburst coincides with the next periastron passage. The current astrometric data still allow a wide range of orbital periods, from \simil25 to \simil100 years, but as new observations accumulate over the coming years, we should be able to fairly accurately predict the onset of the next outburst long before it would occur.

\subsection{Issues}
Here we briefly discuss some issues that are relevant to the current discussion.

\subsubsection{The mass and lifetime of the Sa disk}
\label{sec:Sa_disk_mass}
The amount of material that we can reasonable assume to be present in the Sa disk is at most a few times 10$^{-3}$\,\Msun. If the average accretion rate during an outburst is 5\,$\times$\,10$^{-6}$\,\msunyr \ and an outburst lasts for 10 years, we can have only on the order of 10$^{2}$ these events before the disk is all gone, unless the Sa disk can be supplied with fresh material. If the orbital period of Sa-Sb is 50\,yrs, this means the current situation can exist for only $\lesssim$5000\,yrs. Possibly the Sa-Sb pair became a close binary only within the last several thousand years due to gravitational interaction with a third object. It may also be possible that the Sa disk is still supplied with fresh material coming in at high latitudes with respect to the orbital plane as the Sa-Sb pair moves through ambient cloud material. Because the disks of Sa and Sb appear strongly misaligned \citep{2005ApJ...628..832D,2009A&A...502..623R}, the Sa disk and the Sa-Sb orbit are misaligned, and cloud material is present close to the Sa-Sb system \citep[\sek\,\ref{sec:naco_results},][]{2007AJ....134..359H,2008ApJ...676..472B,2008A&A...488..235G}, the notion that cloud material is still being freshly supplied to the Sa disk is realistic.

\subsubsection{A note on the Sb disk}
\label{sec:Sb_disk}
Like the primary star Sa, the low-mass companion Sb also shows a substantial infrared excess, indicating that some circumstellar material is still present, presumably also in a small circumstellar disk. As our calculations show, Sb will strongly disturb the Sa disk during periastron passage. Conversely, the gravitational pull of Sa will affect the disk around Sb even more strongly, and accretion onto Sb should also be enhanced during periastron passage. Therefore one may wonder why the observed brightness variations are much smaller in Sb (maximum amplitude \simil0.4\,mag $RMS$$=$0.2\,mag) than in Sa (amplitude $\gtrsim$3\,mag). 

\cite{2002ApJ...568..771D,2005ApJ...628..832D} measured the Br\,$\gamma$ line flux of Sb to be 7.2$\times$10$^{-17}$\,Wm$^{-2}$ on 2000 November 19, and 9.4$\times$10$^{-17}$\,Wm$^{-2}$ on 2003 December 12, with corresponding K-band continuum magnitudes of 8.77 and 8.50, respectively. the spectral type of Sb corresponds to $T_{\rm{eff}}$\simil3800\,K \citep{2005ApJ...628..832D}, and assuming an age of 2\,Myr this yields a mass of approximately 0.6\,\Msun \ and a luminosity of \simil0.45\,\Lsun \ by comparison to the pre-main sequence evolutionary tracks of \cite{2008ApJS..178...89D}, which seem reasonable. The corresponding stellar radius of Sb is $R_*$$=$1.55\,\Rsun. Assuming a foreground extinction of $A_{\rm{V}}$$=$15\,mag, a standard extinction law ($A_{\rm{K}}$=0.112\,$A_{\rm{V}}$), the relation between Br\,$\gamma$ luminosity and accretion luminosity given in equation~3 of \cite{2004AJ....128.1294C}, and using the approximate relation $L_{\rm{acc}}$$=$G$M_*$$\dot M_{\rm{acc}}$/$R_*$, one derives mass accretion rates of 3.5$\times$10$^{-8}$\,\msunyr \ and 4.4$\times$10$^{-8}$\,\msunyr \ for the respective epochs. The corresponding accretion luminosities are approximately equal to the stellar luminosity. Therefore, variable accretion is a viable mechanism for the observed modest near-infrared variations in Sb as well. How variations in the accretion luminosity translate to variations in the near-infrared continuum excess emission is somewhat dependent on the disk structure: the extra luminosity is released in the UV which cannot be observed due to the high extinction, and we only see the reprocessed radiation from the hot inner disk regions.

We note, however, that the existing data also perfectly agree with a
time variable extinction towards Sb. Between the aforementioned epochs of Br\,$\gamma$ spectroscopy by Duch\^ene et al. the extinction would have decreased by $\Delta$$A_{\rm{V}}$\simil2.4\,mag, assuming a standard extinction law.

\subsubsection{Mid-infrared brightness during the peak of the outburst}
We are not aware of mid-IR measurements performe between roughly 1998 January and 2001 November, but we argue Sa must have been very bright during that period, peaking at \simil40\,Jy at 12.8\,\mum. Should any mid-IR observations have been performed during this period, they are highly relevant to the current discussion. The variable extinction scenario predicts a lower peak-brightness in mid-IR than the variable accretion scenario. However, only if mid-IR observations have been performed (near-) simultaneously with the existing K-band data they may provide supportive evidenced for either scenario. In case of a substantial time lag the potential short-term variations will inhibit drawing firm conclusions.

\section{Summary}
\label{sec:summary}

The southern infrared companion of T\,Tau is a close binary system of which the more massive member, T\,Tau\,Sa, shows strong long-term photometric variability, with an amplitude of $\gtrsim$3\,mag at 2.2\,\mum \ and \simil2\,mag in the 10\,\mum \ region on timescales of years. Short-term variability of \simil1\,mag within one week has been detected at 2.2\,\mum. The physical mechanism driving the observed flux variations has been under debate, the originally proposed intrinsic luminosity variations due to a variable accretion rate have later been challenged in favor of apparent brightness fluctuations due to variable foreground extinction.

We revisited the nature of the photometric variability of T\,Tau\,Sa using new observations in the mid-infrared. In February 2008 we detected a very fast increase of the mid-IR brightness of T\,Tau\,S which can be attributed to Sa: $+$26$\pm$2\% at 12.8\,$\mu$m in 3.94 days. Using simple geometric arguments and basic physic laws we can exclude that this \emph{short-term} variability is due to time-variable extinction. We show that the fast brightness fluctuations must instead reflect changes in intrinsic luminosity, and attribute these to variable accretion. The key aspect of our argument is the short timescale of the variations in combination with the long observing wavelength.

We show that variable accretion can plausibly explain also the much larger \emph{long-term} variability and its associated near-infrared ``bluer when brighter'' color-magnitude behavior, by creating a 2D radiative transfer disk model of the Sa disk that qualitatively reproduces the observed properties. However, the hitherto observed long-term variability can also be explained within the variable foreground-extinction scenario. A combination of both mechanisms, variable accretion and variable foreground extinction, may be required to explain the collective photometric variability of T\,Tau\,Sa.

In view of the tentative scenario that both the short- and long-term variability of Sa are caused by variable accretion, we reviewed the infrared light curve in relation to the parameters of the Sa-Sb orbit. We propose that the dramatic brightening events in the early and late 1990s and the interspersed episodes of irregular flux variations were an accretion outburst, induced by gravitational perturbation of the Sa disk during the periastron passage of Sb, with closest approach in \simil1995. The induced maximum accretion rate is on the order of 10$^{-5}$\,\msunyr. While we do not model the physics of this outburst in detail, we make a clear observational prediction: if this scheme is correct, Sa should return to ``quiescence'' some time in the near future, and remain in a low state until the next periastron passage of Sb.

\begin{acknowledgements}
We thank the ESO staff for executing the VISIR observations in service mode. RvB thanks Kees Dullemond for many insightful discusisons. An anonymous referee is gratefully acknowledged for a detailed and constructively critical report, from which the manuscript greatly benefited.
\end{acknowledgements}

\bibliographystyle{aa}
\bibliography{references.bib}

\begin{thebibliography}{57}
\expandafter\ifx\csname natexlab\endcsname\relax\def\natexlab#1{#1}\fi

\bibitem[{{Akeson} {et~al.}(1998){Akeson}, {Koerner}, \&
  {Jensen}}]{1998ApJ...505..358A}
{Akeson}, R.~L., {Koerner}, D.~W., \& {Jensen}, E.~L.~N. 1998, \apj, 505, 358

\bibitem[{{Ambartsumian}(1947)}]{ambartsumian1947}
{Ambartsumian}, V.~A. 1947, Stellar Evolution and Astrophysics (Erevan: Acad.
  Sci. Armenian S. S. R.

\bibitem[{{Ambartsumian}(1949)}]{ambartsumian1949}
{Ambartsumian}, V.~A. 1949, AZh, 26, 3

\bibitem[{{Artymowicz}(1993)}]{1993ApJ...419..166A}
{Artymowicz}, P. 1993, \apj, 419, 166

\bibitem[{{Beck} {et~al.}(2008){Beck}, {McGregor}, {Takami}, \&
  {Pyo}}]{2008ApJ...676..472B}
{Beck}, T.~L., {McGregor}, P.~J., {Takami}, M., \& {Pyo}, T.-S. 2008, \apj,
  676, 472

\bibitem[{{Beck} {et~al.}(2001){Beck}, {Prato}, \&
  {Simon}}]{2001ApJ...551.1031B}
{Beck}, T.~L., {Prato}, L., \& {Simon}, M. 2001, \apj, 551, 1031

\bibitem[{{Beck} {et~al.}(2004){Beck}, {Schaefer}, {Simon}, {Prato}, {Stoesz},
  \& {Howell}}]{2004ApJ...614..235B}
{Beck}, T.~L., {Schaefer}, G.~H., {Simon}, M., {et~al.} 2004, \apj, 614, 235

\bibitem[{{Beck} \& {Simon}(2001)}]{2001AJ....122..413B}
{Beck}, T.~L. \& {Simon}, M. 2001, \aj, 122, 413

\bibitem[{{Beckwith} {et~al.}(1984){Beckwith}, {Skrutskie}, {Zuckerman}, \&
  {Dyck}}]{1984ApJ...287..793B}
{Beckwith}, S., {Skrutskie}, M.~F., {Zuckerman}, B., \& {Dyck}, H.~M. 1984,
  \apj, 287, 793

\bibitem[{{Brandner} {et~al.}(2002){Brandner}, {Rousset}, {Lenzen}, {Hubin},
  {Lacombe}, {Hofmann}, {Moorwood}, {Lagrange}, {Gendron}, {Hartung}, {Puget},
  {Ageorges}, {Biereichel}, {Bouy}, {Charton}, {Dumont}, {Fusco}, {Jung},
  {Lehnert}, {Lizon}, {Monnet}, {Mouillet}, {Moutou}, {Rabaud}, {R{\"o}hrle},
  {Skole}, {Spyromilio}, {Storz}, {Tacconi-Garman}, \&
  {Zins}}]{2002Msngr.107....1B}
{Brandner}, W., {Rousset}, G., {Lenzen}, R., {et~al.} 2002, The Messenger, 107,
  1

\bibitem[{{Calvet} {et~al.}(1997){Calvet}, {Hartmann}, \&
  {Strom}}]{1997ApJ...481..912C}
{Calvet}, N., {Hartmann}, L., \& {Strom}, S.~E. 1997, \apj, 481, 912

\bibitem[{{Calvet} {et~al.}(2004){Calvet}, {Muzerolle}, {Brice{\~n}o},
  {Hern{\'a}ndez}, {Hartmann}, {Saucedo}, \& {Gordon}}]{2004AJ....128.1294C}
{Calvet}, N., {Muzerolle}, J., {Brice{\~n}o}, C., {et~al.} 2004, \aj, 128, 1294

\bibitem[{{Cohen} {et~al.}(1999){Cohen}, {Walker}, {Carter}, {Hammersley},
  {Kidger}, \& {Noguchi}}]{1999AJ....117.1864C}
{Cohen}, M., {Walker}, R.~G., {Carter}, B., {et~al.} 1999, \aj, 117, 1864

\bibitem[{{Diolaiti} {et~al.}(2000){Diolaiti}, {Bendinelli}, {Bonaccini},
  {Close}, {Currie}, \& {Parmeggiani}}]{2000A&AS..147..335D}
{Diolaiti}, E., {Bendinelli}, O., {Bonaccini}, D., {et~al.} 2000, \aaps, 147,
  335

\bibitem[{{Dotter} {et~al.}(2008){Dotter}, {Chaboyer}, {Jevremovi{\'c}},
  {Kostov}, {Baron}, \& {Ferguson}}]{2008ApJS..178...89D}
{Dotter}, A., {Chaboyer}, B., {Jevremovi{\'c}}, D., {et~al.} 2008, \apjs, 178,
  89

\bibitem[{{Duch{\^e}ne} {et~al.}(2006){Duch{\^e}ne}, {Beust}, {Adjali},
  {Konopacky}, \& {Ghez}}]{2006A&A...457L...9D}
{Duch{\^e}ne}, G., {Beust}, H., {Adjali}, F., {Konopacky}, Q.~M., \& {Ghez},
  A.~M. 2006, \aap, 457, L9

\bibitem[{{Duch{\^e}ne} {et~al.}(2002){Duch{\^e}ne}, {Ghez}, \&
  {McCabe}}]{2002ApJ...568..771D}
{Duch{\^e}ne}, G., {Ghez}, A.~M., \& {McCabe}, C. 2002, \apj, 568, 771

\bibitem[{{Duch{\^e}ne} {et~al.}(2005){Duch{\^e}ne}, {Ghez}, {McCabe}, \&
  {Ceccarelli}}]{2005ApJ...628..832D}
{Duch{\^e}ne}, G., {Ghez}, A.~M., {McCabe}, C., \& {Ceccarelli}, C. 2005, \apj,
  628, 832

\bibitem[{{Dullemond} \& {Dominik}(2004)}]{2004A&A...417..159D}
{Dullemond}, C.~P. \& {Dominik}, C. 2004, \aap, 417, 159

\bibitem[{{Dyck} {et~al.}(1982){Dyck}, {Simon}, \&
  {Zuckerman}}]{1982ApJ...255L.103D}
{Dyck}, H.~M., {Simon}, T., \& {Zuckerman}, B. 1982, \apjl, 255, L103

\bibitem[{{Furlan} {et~al.}(2003){Furlan}, {Forrest}, {Watson}, {Uchida},
  {Brandl}, {Keller}, \& {Herter}}]{2003ApJ...596L..87F}
{Furlan}, E., {Forrest}, W.~J., {Watson}, D.~M., {et~al.} 2003, \apjl, 596, L87

\bibitem[{{Ghez} {et~al.}(1991){Ghez}, {Neugebauer}, {Gorham}, {Haniff},
  {Kulkarni}, {Matthews}, {Koresko}, \& {Beckwith}}]{1991AJ....102.2066G}
{Ghez}, A.~M., {Neugebauer}, G., {Gorham}, P.~W., {et~al.} 1991, \aj, 102, 2066

\bibitem[{{Gustafsson} {et~al.}(2008){Gustafsson}, {Labadie}, {Herbst}, \&
  {Kasper}}]{2008A&A...488..235G}
{Gustafsson}, M., {Labadie}, L., {Herbst}, T.~M., \& {Kasper}, M. 2008, \aap,
  488, 235

\bibitem[{{Hatzes} {et~al.}(2003){Hatzes}, {Cochran}, {Endl}, {McArthur},
  {Paulson}, {Walker}, {Campbell}, \& {Yang}}]{2003ApJ...599.1383H}
{Hatzes}, A.~P., {Cochran}, W.~D., {Endl}, M., {et~al.} 2003, \apj, 599, 1383

\bibitem[{{Herbst} {et~al.}(2007){Herbst}, {Hartung}, {Kasper}, {Leinert}, \&
  {Ratzka}}]{2007AJ....134..359H}
{Herbst}, T.~M., {Hartung}, M., {Kasper}, M.~E., {Leinert}, C., \& {Ratzka}, T.
  2007, \aj, 134, 359

\bibitem[{{Herbst} {et~al.}(1997){Herbst}, {Robberto}, \&
  {Beckwith}}]{1997AJ....114..744H}
{Herbst}, T.~M., {Robberto}, M., \& {Beckwith}, S.~V.~W. 1997, \aj, 114, 744

\bibitem[{{Herbst} {et~al.}(1994){Herbst}, {Herbst}, {Grossman}, \&
  {Weinstein}}]{1994AJ....108.1906H}
{Herbst}, W., {Herbst}, D.~K., {Grossman}, E.~J., \& {Weinstein}, D. 1994, \aj,
  108, 1906

\bibitem[{{Hogerheijde} {et~al.}(1997){Hogerheijde}, {van Langevelde}, {Mundy},
  {Blake}, \& {van Dishoeck}}]{1997ApJ...490L..99H}
{Hogerheijde}, M.~R., {van Langevelde}, H.~J., {Mundy}, L.~G., {Blake}, G.~A.,
  \& {van Dishoeck}, E.~F. 1997, \apjl, 490, L99+

\bibitem[{{Joy}(1945)}]{1945ApJ...102..168J}
{Joy}, A.~H. 1945, \apj, 102, 168

\bibitem[{{Kasper} {et~al.}(2002){Kasper}, {Feldt}, {Herbst}, {Hippler}, {Ott},
  \& {Tacconi-Garman}}]{2002ApJ...568..267K}
{Kasper}, M.~E., {Feldt}, M., {Herbst}, T.~M., {et~al.} 2002, \apj, 568, 267

\bibitem[{{Kley} \& {Nelson}(2008)}]{2008A&A...486..617K}
{Kley}, W. \& {Nelson}, R.~P. 2008, \aap, 486, 617

\bibitem[{{Kobayashi} {et~al.}(1994){Kobayashi}, {Nagata}, {Hodapp}, \&
  {Hora}}]{1994PASJ...46L.183K}
{Kobayashi}, N., {Nagata}, T., {Hodapp}, K.-W., \& {Hora}, J.~L. 1994, \pasj,
  46, L183

\bibitem[{{K{\"o}hler}(2008)}]{2008JPhCS.131a2028K}
{K{\"o}hler}, R. 2008, Journal of Physics Conference Series, 131, 012028

\bibitem[{{K{\"o}hler} {et~al.}(2008){K{\"o}hler}, {Ratzka}, {Herbst}, \&
  {Kasper}}]{2008A&A...482..929K}
{K{\"o}hler}, R., {Ratzka}, T., {Herbst}, T.~M., \& {Kasper}, M. 2008, \aap,
  482, 929

\bibitem[{{Koresko}(2000)}]{2000ApJ...531L.147K}
{Koresko}, C.~D. 2000, \apjl, 531, L147

\bibitem[{{Kuhi}(1974)}]{1974A&AS...15...47K}
{Kuhi}, L.~V. 1974, \aaps, 15, 47

\bibitem[{{Lagage} {et~al.}(2004){Lagage}, {Pel}, {Authier}, {Belorgey},
  {Claret}, {Doucet}, {Dubreuil}, {Durand}, {Elswijk}, {Girardot}, {K{\"a}ufl},
  {Kroes}, {Lortholary}, {Lussignol}, {Marchesi}, {Pantin}, {Peletier},
  {Pirard}, {Pragt}, {Rio}, {Schoenmaker}, {Siebenmorgen}, {Silber}, {Smette},
  {Sterzik}, \& {Veyssiere}}]{2004Msngr.117...12L}
{Lagage}, P.~O., {Pel}, J.~W., {Authier}, M., {et~al.} 2004, The Messenger,
  117, 12

\bibitem[{{Larson}(1981)}]{1981MNRAS.194..809L}
{Larson}, R.~B. 1981, \mnras, 194, 809

\bibitem[{{Loinard} {et~al.}(2007){Loinard}, {Torres}, {Mioduszewski},
  {Rodr{\'{\i}}guez}, {Gonz{\'a}lez-L{\'o}pezlira}, {Lachaume}, {V{\'a}zquez},
  \& {Gonz{\'a}lez}}]{2007ApJ...671..546L}
{Loinard}, L., {Torres}, R.~M., {Mioduszewski}, A.~J., {et~al.} 2007, \apj,
  671, 546

\bibitem[{{Lord}(1992)}]{lord92}
{Lord}, S.~D. 1992, NASA Technical Memorandum 103957

\bibitem[{{Lozinskii}(1949)}]{lozinskii_1949}
{Lozinskii}, A.~M. 1949, Perem. Zvezdy, 7, 76

\bibitem[{{Maihara} \& {Kataza}(1991)}]{1991A&A...249..392M}
{Maihara}, T. \& {Kataza}, H. 1991, \aap, 249, 392

\bibitem[{{Mel'nikov} \& {Grankin}(2005)}]{2005AstL...31..427M}
{Mel'nikov}, S.~Y. \& {Grankin}, K.~N. 2005, Astronomy Letters, 31, 427

\bibitem[{{Przygodda} {et~al.}(2003){Przygodda}, {van Boekel},
  {{\`A}brah{\`a}m}, {Melnikov}, {Waters}, \& {Leinert}}]{2003A&A...412L..43P}
{Przygodda}, F., {van Boekel}, R., {{\`A}brah{\`a}m}, P., {et~al.} 2003, \aap,
  412, L43

\bibitem[{{Ratzka} {et~al.}(2009){Ratzka}, {Schegerer}, {Leinert},
  {{\'A}brah{\'a}m}, {Henning}, {Herbst}, {K{\"o}hler}, {Wolf}, \&
  {Zinnecker}}]{2009A&A...502..623R}
{Ratzka}, T., {Schegerer}, A.~A., {Leinert}, C., {et~al.} 2009, \aap, 502, 623

\bibitem[{{Reimann} {et~al.}(1998){Reimann}, {Weinert}, \&
  {Wagner}}]{1998SPIE.3354..865R}
{Reimann}, H., {Weinert}, U., \& {Wagner}, S. 1998, in Proc. SPIE Vol. 3354, p.
  865-876, Infrared Astronomical Instrumentation, Albert M. Fowler; Ed., Vol.
  3354, 865--876

\bibitem[{{Rieke} \& {Lebofsky}(1985)}]{1985ApJ...288..618R}
{Rieke}, G.~H. \& {Lebofsky}, M.~J. 1985, \apj, 288, 618

\bibitem[{{Roddier} {et~al.}(2000){Roddier}, {Roddier}, {Brandner},
  {Charissoux}, {V{\'e}ran}, \& {Courbin}}]{2000IAUS..200P..60R}
{Roddier}, F., {Roddier}, C., {Brandner}, W., {et~al.} 2000, in IAU Symposium,
  Vol. 200, IAU Symposium, 60P--+

\bibitem[{{Shakura} \& {Sunyaev}(1973)}]{1973A&A....24..337S}
{Shakura}, N.~I. \& {Sunyaev}, R.~A. 1973, \aap, 24, 337

\bibitem[{{Simon} {et~al.}(1996){Simon}, {Longmore}, {Shure}, \&
  {Smillie}}]{1996ApJ...456L..41S}
{Simon}, M., {Longmore}, A.~J., {Shure}, M.~A., \& {Smillie}, A. 1996, \apjl,
  456, L41+

\bibitem[{{Skemer} {et~al.}(2008){Skemer}, {Close}, {Hinz}, {Hoffmann},
  {Kenworthy}, \& {Miller}}]{2008ApJ...676.1082S}
{Skemer}, A.~J., {Close}, L.~M., {Hinz}, P.~M., {et~al.} 2008, \apj, 676, 1082

\bibitem[{{Stapelfeldt} {et~al.}(1998){Stapelfeldt}, {Burrows}, {Krist},
  {Watson}, {Ballester}, {Clarke}, {Crisp}, {Evans}, {Gallagher}, {Griffiths},
  {Hester}, {Hoessel}, {Holtzman}, {Mould}, {Scowen}, {Trauger}, \&
  {Westphal}}]{1998ApJ...508..736S}
{Stapelfeldt}, K.~R., {Burrows}, C.~J., {Krist}, J.~E., {et~al.} 1998, \apj,
  508, 736

\bibitem[{{Tessier} {et~al.}(1994){Tessier}, {Bouvier}, \&
  {Lacombe}}]{1994A&A...283..827T}
{Tessier}, E., {Bouvier}, J., \& {Lacombe}, F. 1994, \aap, 283, 827

\bibitem[{{van Boekel} {et~al.}(2005){van Boekel}, {Dullemond}, \&
  {Dominik}}]{2005A&A...441..563V}
{van Boekel}, R., {Dullemond}, C.~P., \& {Dominik}, C. 2005, \aap, 441, 563

\bibitem[{{van Boekel} {et~al.}(2009){van Boekel}, {G{\"u}del}, {Henning},
  {Lahuis}, \& {Pantin}}]{2009A&A...497..137V}
{van Boekel}, R., {G{\"u}del}, M., {Henning}, T., {Lahuis}, F., \& {Pantin}, E.
  2009, \aap, 497, 137

\bibitem[{{van den Ancker} {et~al.}(1999){van den Ancker}, {Wesselius},
  {Tielens}, {van Dishoeck}, \& {Spinoglio}}]{1999A&A...348..877V}
{van den Ancker}, M.~E., {Wesselius}, P.~R., {Tielens}, A.~G.~G.~M., {van
  Dishoeck}, E.~F., \& {Spinoglio}, L. 1999, \aap, 348, 877

\bibitem[{{White} \& {Ghez}(2001)}]{2001ApJ...556..265W}
{White}, R.~J. \& {Ghez}, A.~M. 2001, \apj, 556, 265

\end{thebibliography}

\end{document}